%% file: ms.tex
\documentclass[10pt,apjl]{emulateapj}
\usepackage[dvips]{epsfig}
\usepackage[dvips]{color}
\usepackage[usenames,dvipsnames]{xcolor}

\begin{document}
\newcommand{\boo}   {Bo\"{o}tes~I}
\newcommand{\boos}  {Boo-1137}
\newcommand{\kms}   {\rm km~s$^{-1}$} 
\newcommand{\teff}  {$T_{\rm eff}$} 
\newcommand{\logg}  {$\log g$} 
\newcommand{\loggf} {log~$gf$} 
\newcommand{\bvz}   {$(B-V)_{0}$} 
\newcommand{\feh}   {[Fe/H]}
\newcommand{\vxi}   {$\xi_t$}
\newcommand{\uf}    {ultra-faint}
\newcommand{\msun} {\rm M$_\odot$}
\newcommand{\ltsima} {$\; \buildrel < \over \sim \;$}
\newcommand{\simlt} {\lower.5ex\hbox{\ltsima}}
\newcommand{\gtsima} {$\; \buildrel > \over \sim \;$}
\newcommand{\simgt} {\lower.5ex\hbox{\gtsima}}

\shorttitle{THE CHEMICAL ENRICHMENT OF BO\"{O}TES~I}
\shortauthors{Gilmore et al.}

\title{ELEMENTAL ABUNDANCES AND THEIR IMPLICATIONS FOR THE CHEMICAL
  ENRICHMENT OF THE BO\"{O}TES~I ULTRA-FAINT GALAXY\footnote{Based on
    observations 
    collected at the European Southern Observatory, Paranal, Chile
    (Proposal P82.182.B-0372, PI: G. Gilmore)}}

\author{
GERARD GILMORE\altaffilmark{1},
JOHN E. NORRIS\altaffilmark{2},
LORENZO MONACO\altaffilmark{3},
DAVID YONG\altaffilmark{2},
ROSEMARY F.G. WYSE\altaffilmark{4}, AND
D. GEISLER\altaffilmark{5}
}

\altaffiltext{1}{Institute of Astronomy, University of Cambridge,
  Madingley Road, Cambridge CB3 0HA, UK; gil@ast.cam.ac.uk}

\altaffiltext{2}{Research School of Astronomy and Astrophysics, The
  Australian National University, Weston, ACT 2611, Australia;
  jen@mso.anu.edu.au, yong@mso.anu.edu.au}

\altaffiltext{3}{European Southern Observatory, Alonso de Cordova
  3107, Casilla 19001, Santiago 19, Chile; lmonaco@eso.org }

\altaffiltext{4}{The Johns Hopkins University, Department of Physics
  \& Astronomy, 3900 N.~Charles Street, Baltimore, MD 21218, USA;
  wyse@pha.jhu.edu}

\altaffiltext{5}{Departamento de Astronomia, Universidad de
  Concepcion, Chile; dgeisler@astro-udec.cl}

\begin{abstract}

We present a double-blind analysis of high-dispersion spectra of seven
red giant members of the {\boo} {\uf} dwarf spheroidal galaxy,
complemented with re-analysis of a similar spectrum of an eighth
member star.  The stars cover [Fe/H] from --3.7 to --1.9, and include
a CEMP-no star with [Fe/H] = --3.33.  We conclude from our chemical
abundance data that {\boo} has evolved as a self-enriching
star-forming system, from essentially primordial initial
abundances. This allows us uniquely to investigate the place of
CEMP-no stars in a chemically evolving system, in addition to limiting
the timescale of star formation.  The elemental abundances are
formally consistent with a halo-like distribution, with enhanced mean
[$\alpha$/Fe] and small scatter about the mean. This is in accord with
the high-mass stellar IMF in this low-stellar-density, low-metallicity
system being indistinguishable from the present-day solar neighborhood
value.  There is a non-significant hint of a decline in [$\alpha$/Fe]
with [Fe/H]; together with the low scatter, this requires low star
formation rates, allowing time for SNe ejecta to be mixed over the
large spatial scales of interest. One star has very high [Ti/Fe], but
we do not confirm a previously published high value of [Mg/Fe] for
another star.  We discuss the existence of CEMP-no stars, and the
absence of any stars with lower CEMP-no enhancements at higher [Fe/H],
a situation which is consistent with knowledge of CEMP-no stars in the
Galactic field.  We show that this observation requires there be two
enrichment paths at very low metallicities: CEMP-no and
``carbon-normal''.

\end{abstract}


\section{INTRODUCTION} \label{Sec:intro}

The {\boo} ultra-faint dwarf spheroidal galaxy was discovered by
\citet{belokurov06}, who reported an absolute magnitude M$_{V,\,total}
= -5.8$ and half-light radius $\sim220$~pc, noting that its magnitude
``makes it one of the faintest galaxies known''.  This discovery has
been followed by a large number of investigations of the spatial,
kinematic, and chemical abundance distributions of {\boo}, all aiming
to provide insight into the formation and evolution of this extremely
low luminosity galaxy, and into what it has to tell us about
conditions at the earliest times.  Photometric studies indicate an
exclusively old stellar population. We refer the reader to the works
of \citet{belokurov06}, \citet{dallora06}, \citet{fellhauer08},
\citet{feltzing09}, \citet{koposov11}, \citet{lai11}, \citet{martin07,
  martin08}, \citet{munoz06}, \citet{norris08, norris10a, norris10b},
and \citet{okamoto12} for details of progress to date.  Two
fundamental results have emerged.  The first is that this low
luminosity galaxy is dark-matter dominated, with a mass-to-light ratio
within the half-light radius estimated to lie in the range 120 $< M/L
<$ 1700 (\citealt{martin07, wolf10, koposov11}).  The kinematics may
be complex: \citet{koposov11} tentatively identify two kinematically
distinguishable components, and speculate that this ``reflects the
distribution of velocity anisotropy in {\boo}, which is a measure of
its formation processes.''  The second result is that there is also a
large dispersion in chemical abundances within the system, indicative
of self-enrichment: for iron the range is $-3.7 <$ [Fe/H]
\footnote{Here we adopt [Fe/H] = log(N$_{\rm Fe}$/N$_{\rm H})_\star$
  $-$ log(N$_{\rm Fe}$/N$_{\rm H})_\odot$ (where N$_{\rm X}$ is the
  number of atoms of element X); [X/Fe] = log(N$_{\rm X}$/N$_{\rm
    Fe})_\star$ $-$ log(N$_{\rm X}$/N$_{\rm Fe})_\odot$; and
  log\,$\epsilon$(X) = log(N$_{\rm X}$/N$_{\rm H}$) + 12.0} $<-1.9$;
there is a wide range in the relative abundance of carbon: --0.8 $<$
[C/Fe] $<$ +2.2 \citep{norris10b,lai11}; and \citet{feltzing09}
describe one object with an atypically high value of [Mg/Ca] $\sim$
+0.7.

{\boo} is one of the brighter of some 15 newly recognized ultra-faint
dwarf galaxy satellites of the Milky Way, which are the subject of
considerable current activity, driven by their potential to provide an
understanding of fundamental questions on the formation and evolution
of galaxies (see, e.g., \citealt{gilmore07} and references therein).
For example, why are there considerably fewer dwarf galaxy satellites
of the Milky Way than predicted by the $\Lambda$CDM paradigm; what is
the connection between their bright and dark matter; when and where
did their baryonic component form; and what has driven their chemical
abundance inhomogeneities?

The present paper is the fourth in a series aimed at understanding the
chemical abundance characteristics of {\boo}, and their implications
for the chemical enrichment of the system and the manner in which it
formed.  The first \citep{norris08} reported abundances for 16
radial-velocity members based on medium-resolution spectra (R
$\sim5000$): we found an abundance range of
$\Delta$[Fe/H]$\sim2.0$~dex, with one star ({\boos}) having [Fe/H]
$\sim -3.4$.  The second \citep{norris10a} confirmed the extremely
metal-poor nature of {\boos} based on high-resolution (R
$\sim40,000$), high-$S/N$ ($S/N \sim20 - 90$), spectroscopy: this star
has [Fe/H] = --3.7, and relative-abundance ratios for some 15
additional elements that are comparable to those of extremely
metal-poor stars of similar [Fe/H] in the Galactic halo.  In the third
paper \citep{norris10b}, we determined carbon abundances ([C/Fe]) for
the 16 radial-velocity members reported in the first paper, together
with preliminary values of [Fe/H] for seven of these stars for which
we had obtained high resolution (R~{$\sim45000$}) spectroscopy: our
conclusion was that the abundance dispersion was real, and the
distribution of the carbon abundances in these red giants was not
unlike that of the Galactic halo.

The purpose of this paper is to report the data of the seven
high-resolution spectra noted above, to present abundance measurements
for 14 elemental species, and to consider both systematic and random
uncertainties in our results.  In Section~\ref{Sec:obsdat} we describe
the observational material and the measurement of line strengths and
radial velocities, while in Sections~\ref{Sec:analysis} and
\ref{Sec:relabund} we analyze these to produce and present chemical
abundances.  As part of our measurement and abundance determination,
we adopt a double-blind methodology that employs two distinct analyses
of the dataset in order to permit us to obtain an independent
assessment of the errors associated with our results.  Finally, in
Section~\ref{Sec:discussion} we discuss the implications of our
results for the chemical evolution of {\boo} and chemical enrichment
at the earliest times.

\subsection{Double-blind Analysis}

Detection of a range of abundances in an ultra-faint dSph galaxy, and
especially detection of either or both of a real range in elemental
abundance ratios at a given iron abundance, or a trend in elemental
abundance ratios as a function of iron abundance, provides constraints
on the rate of star-formation and associated self-enrichment, and the
efficiency, timescale, and length scale of mixing in the interstellar
medium at very early times. Hence, understanding both systematic and
random measuring errors is an essential aspect of an analysis. Before
proceeding to determine abundances from these spectra of stars in
{\boo}, and in an effort to obtain an external estimate of the
abundance accuracy that independent researchers might achieve from
spectra of the quality we have available, the decision was taken to
perform two independent analyses. We refer the reader to
\citet{bensby09} for an earlier example of this type of approach.  It
was agreed that J.E.N. and D.Y. (working together, and hereafter
referred to as NY) would perform one analysis, while D.G. and
L.M. (also working together, and referred to as GM) would perform the
other.  There would be no correspondence or discussion between NY and
GM in this first phase of the project. Both groups would use their
standard approaches, line lists, etc, consistent with their previous
published studies.  The reader will see this reflected in
Sections~\ref{Sec:radvel} -- \ref{Sec:errors}, which describe the
measurement and analysis of our spectra to produce radial velocities,
stellar atmospheric parameters, and chemical abundances.
   
\section{HIGH-RESOLUTION SPECTROSCOPY}\label{Sec:obsdat}

\subsection{Observational Data}\label{Sec:obsspec}

High-resolution, moderate-$S/N$, spectra were obtained of seven {\boo} red
giants as part of a larger program to investigate the kinematics and
chemical abundances of the {\boo} system.  During 2009 February --
March, data were obtained with the FLAMES spectrograph of the 8.2 m
Kueyen (VLT/UT2) telescope at Cerro Paranal, Chile.  We used FLAMES in
UVES-Fiber mode \citep{pasquini02}: 130 fibers fed the
medium-resolution Giraffe spectrograph, while eight additional fibers
led to the high-resolution Ultraviolet-Visual Echelle Spectrograph
(UVES).  We refer the reader to \citet {koposov11} for the kinematic
analysis of the medium-resolution data: the UVES spectra are the
subject of the present work.  Of the eight UVES fibers, seven were
allocated to the {\boo} members and one to a nearby sky position to
permit background measurement. 22 useful individual exposures were
obtained in Service Mode, most of duration 46\,min, leading to an
effective total integration time of 17.2 hrs.  The spectra were
obtained using the 580nm setting and cover the wavelength ranges 4800
-- 5750\,{\AA} and 5840 -- 6800\,{\AA}, and have resolving power R =
47,000.

Details of the seven program stars are presented in
Table~\ref{Tab:program}.  Six of them were taken from the sample of
\citet{norris08}, who used medium-resolution spectra to obtain initial
estimates of [Fe/H].  The seventh is from Table 1 of \citet{martin07},
where it is the sixty-third entry.  In what follows we shall refer to
this object as Boo-119, consistent with the identification system
adopted by \citet{norris08}.  In the present Table~\ref{Tab:program},
columns (1) -- (2) present the identifications of \citet{norris08},
and of \citet{martin07} or \citet{lai11}, respectively, while columns
(3) -- (4) contain coordinates.  Column (5) -- (6) present SDSS Data
Release 7
(\citealt{abazajian09}\footnote{{http://cas.sdss.org/astrodr7/en/tools/search/}})
$ugriz$ photometry $g_{0}$ and $(g-r)_{0}$ (where a reddening of
E($B-V$) = 0.02 \citep{belokurov06} has been adopted).  Finally,
columns (7) -- (9) contain [Fe/H] values determined from
medium-resolution spectroscopy by \citet[R~$\sim 5000$]{norris08},
\citet[R~$\sim8500$]{martin07}, and \citet[R~$\sim 1800$]{lai11} using
the Ca II K line, the Ca II infrared triplet, and the blue spectral
region, respectively.  All stars have projected distances from the
nominal centre of {\boo} that are well within one half-light radius
and have radial velocities and values of [Fe/H] consistent with
membership of {\boo} (see Sections~\ref{Sec:radvel} and
~\ref{Sec:analysis} below).

\input{tab1}

\begin{figure}[htbp]
\begin{center}
\includegraphics[width=8.5cm,angle=0]{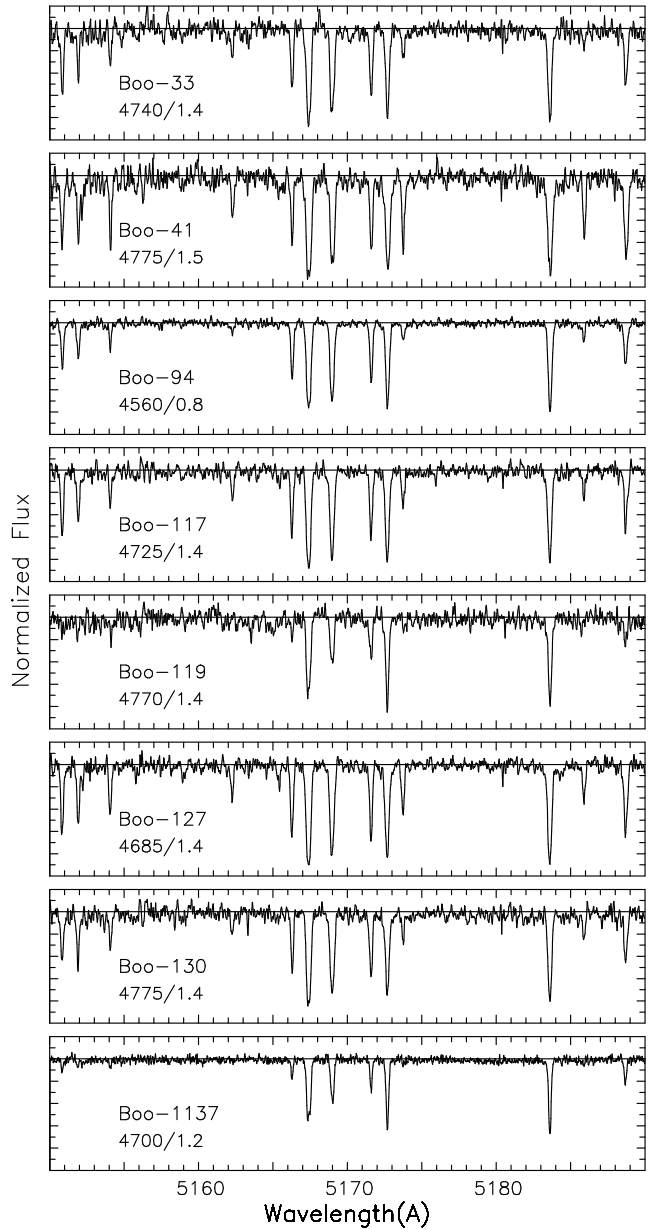}

  \caption{\label{Fig:spectra} Spectra of the {\boo} giants (including
    {\boos} from \citealt{norris10a}) in the region of the Mg b lines.
    Mean adopted values of {\teff}/{\logg} are also presented.}

\end{center}
\end{figure}

The spectra of each individual exposure of the seven {\boo} stars were
reduced with the FLAMES-UVES
pipeline\footnote{{http://www.eso.org/sci/software/pipelines/}}. Examples
of reduced, co-added, and continuum-normalized spectra of the seven
program stars, in the region of the Mg~I~b lines (at 5169.3, 5172.7,
and 5183.6~{\AA}), are shown in Figure~\ref{Fig:spectra}, together
with that of {\boos} from the work of \citet{norris10a}.  Effective
temperatures and surface gravities derived in the analysis below are
also included in the figure.  A large range in line strength among the
sample is clearly evident and as will be demonstrated in the analysis
that follows, a large range in chemical abundance, of order $\Delta
{\rm [X/H]} = 1.8$~dex, is required to explain these differences.

\subsection{Radial Velocities}\label{Sec:radvel}    

Radial velocities were determined independently by NY and GM for the
individual spectra described in Section~\ref{Sec:obsspec}, in order to
confirm membership of {\boo} and to complement the medium-resolution
Giraffe-based investigation of \citet{koposov11}.

\subsubsection{NY analysis}       

Following \citet[Section 2.4]{norris10a}, radial velocities were
measured over the wavelength range 5160 -- 5190\,{\AA}, which contains
the relatively strong Mg~I~b lines, for each of the individual
exposures.  This was achieved by Fourier cross-correlation (using
`scross' in the FIGARO reduction
package\footnote{http://www.aao.gov.au/figaro}) against a synthetic
spectrum having {\teff} = 4700\,K, {\logg} = 1.5, [M/H] = --2.5, and
microturbulent velocity $\xi_t$ = 2~{\kms} (computed with the code
described by \citet{cottrell78}, model atmospheres of
\citet{kurucz93a}, and atomic line data from VALD). The resulting
heliocentric radial velocities are presented in
Table~\ref{Tab:radvels}, where columns (1) -- (4) contain the star
name, radial velocity, (internal) standard error, and the number of
individual spectra that were measured, respectively.  The internal
accuracy is 0.3~{\kms}, small compared with the full velocity spread
of 15~{\kms}.

\input{tab2}

\subsubsection{GM analysis} 

Radial velocities were determined by using the IRAF tasks `fxcor' and
`rvcorrect' to cross-correlate the individual observed
  spectra over the wavelength range 4900 -- 5700\,{\AA} against the
synthetic, high-resolution, spectrum of the \citet{coelho05} model
atmosphere having {\teff} = 4500\,K, {\logg} = 1.5, [Fe/H] = --2.0,
and [$\alpha$/Fe] = +0.4.  The resulting heliocentric velocities,
internal errors, and number of spectra analyzed are shown in columns
(5) -- (7) of Table~\ref{Tab:radvels}.  The mean internal precision of
the velocities is $\sim0.13$~{\kms}, while the full velocity spread is
14 {\kms}.

\subsubsection{Adopted Velocities}

The agreement between the NY and GM velocities is excellent, with the
mean difference between the two determinations being --0.2 {\kms} with
dispersion 0.6 {\kms}.  If one were to exclude Boo-119 (the most
metal-poor star, where template mismatch is anticipated to have
degraded the measurement accuracy) these numbers become --0.1 {\kms}
and 0.1 {\kms}, respectively.  The average values of the NY and GM
velocities for individual stars are given in column (8) of
Table~\ref{Tab:radvels}. These lead to the mean value for the sample
of 100.2 $\pm$ 1.9, with dispersion 5.0 $\pm$ 1.3.  The individual
velocities in column (8) confirm that all of the seven stars have
values consistent with their being members of {\boo}.

\subsubsection{Comparison with the results of \citet{koposov11}}

The present UVES-based mean velocity and dispersion are consistent
with the Giraffe-based values of \citet{koposov11}, who obtained 101.8
$\pm$ 0.7 {\kms} and 4.6~$^{+0.8}_{-0.6}$ {\kms}, respectively, in
their analysis of a sample of some 100 {\boo} members.
\citet{koposov11} reported that their derived stellar radial
velocities could be equally well-fit by models having a single
component with dispersion 4.6~$^{+0.8}_{-0.6}$~{\kms}, or two
kinematically distinct components -- the first with velocity
dispersion 2.4~$^{+0.9}_{-0.5}$~{\kms} which comprises 70\% of the
system, and the second with dispersion ``around'' 9~{\kms} making up
the remaining 30\%. Bearing in mind that our UVES data and the Giraffe
results probe the same projected distances from the centre of {\boo},
we tested whether our radial velocity data are consistent with this
two-component model using Monte Carlo analysis as follows.  Assuming
Gaussian velocity distributions for the two components of
\citet{koposov11}, we drew 10000 samples of seven stars at random and
computed their velocity dispersion.  We found that values greater than
or equal to the observed dispersion of 5 {\kms} are expected
relatively frequently, some $\sim 28\%$ of the time.

\subsection{ Stellar Atmospheric Parameter Determination}

\subsubsection{NY analysis}

In order to perform model atmosphere abundance analyses, one needs the
atmospheric parameters {\teff} and {\logg}.  For the NY analysis our
values are those presented by \citet[Section~5 and
  Table~3]{norris10b}, which we reproduce here in columns (2) and (3)
of Table~\ref{Tab:param}.  While we refer the reader to the earlier
work for details, we note here that (i) temperatures are based on
calibrations of $B-V$ and $griz$ photometry following \citet{norris08}
and
Castelli\footnote{http://wwwuser.oat.ts.astro.it/castelli/colors/sloan.html},
respectively, and (ii) gravities were obtained by comparing the colors
with those of the Yale--Yonsei (YY) Isochrones
(\citealp{demarque04}\footnote{http://www.astro.yale.edu/demarque/yyiso.html})
adopting an age of 12 Gyr, and the assumption that the stars lie on
the red giant branch of the system.  These determinations require
chemical abundance as an input parameter: the medium-resolution [Fe/H]
values of \citet{norris08} and \citet{martin07} (see columns (7) and
(8) of our Table~\ref{Tab:program}, respectively) were used to provide
first estimates of {\teff} and {\logg} and thence model atmosphere
abundances, and the process iterated until self-consistent values were
obtained.

\input{tab3}

\subsubsection{GM analysis}

The stellar atmospheric parameters {\teff} and {\logg} were determined
by GM by comparing the SDSS $gr$ photometry presented in Table~
\ref{Tab:program} with the BaSTI $\alpha$-enhanced isochrones
\citep{pietrinferni06}\footnote{http://193.204.1.62/index.html} in the
($M_{g}$, $g-r$) (absolute magnitude, color) -- plane for an adopted
age of 12 Gyr, a distance modulus of $(m-M)_{0}$ = 19.10
\citep{dallora06}, and a reddening of E($B-V$) = 0.02.  This process
also requires an estimate of metallicity.  Based on an iterative
abundance procedure, GM adopted isochrones having [Fe/H] = --2.6 for
Boo-94 and Boo-119, and --2.1 for the other stars.  GM's adopted
values of {\teff} and {\logg} are presented in columns (6) and (7) of
Table~\ref{Tab:param}.

\subsection{Equivalent Widths}\label{Sec:equivalentwidths} 

\subsubsection{NY measurements}\label{Sec:equivny}

The individual ESO VLT pipeline-reduced spectra were first
cross-correlated to determine relative wavelength shifts between them
in order to compensate for the Earth's motion during the $\sim40$ day
data-taking interval.  After sky-subtracting and shifting the
individual spectra to the rest frame, NY co-added and then
double-binned them into pixels of width 0.028\,{\AA} and 0.034\,{\AA}
(for the shorter and longer wavelength regions described in
Section~\ref{Sec:obsspec}). These spectra were smoothed with a
Gaussian of standard deviation $\sim0.025$~{\AA} to produce final
spectra for the seven program stars. For six of the stars the $S/N$
per 0.03\,{\AA} pixel at 5500~{\AA} was 25 -- 35, while for the
seventh (Boo-94) it was 60.

\begin{figure}[htbp]
\begin{center}
\includegraphics[width=8.5cm,angle=0]{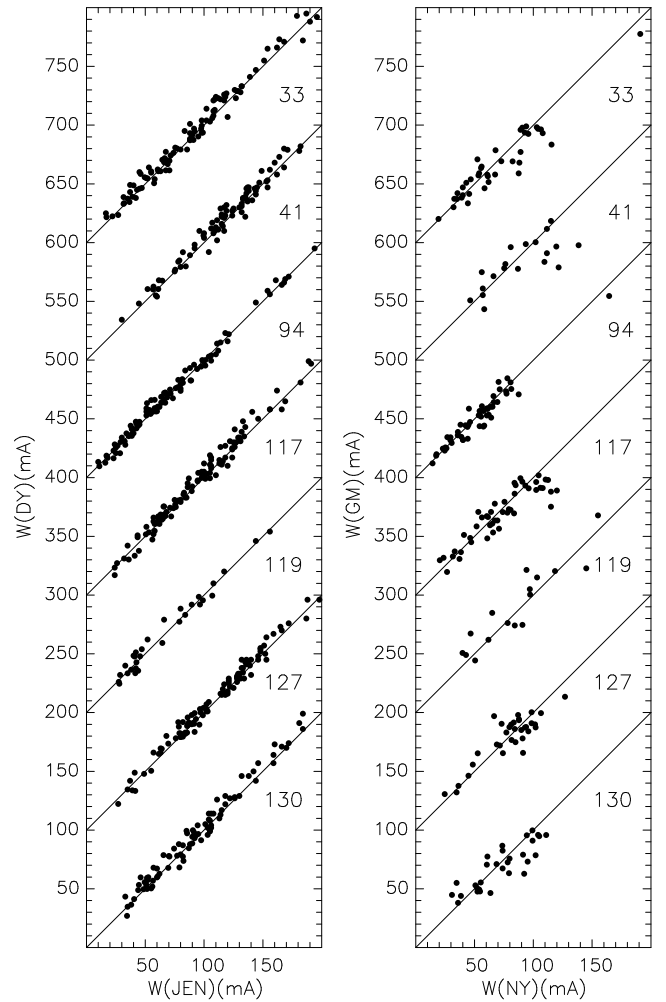}

  \caption{\label{Fig:ew} Comparison of equivalent widths of (left)
    J.E.N. and D.Y., and (right) NY and GM for the seven {\boo}
    giants, as described in the text.  (For convenience of
    presentation, individual stars have been offset by multiples of
    100\,m{\AA} in the vertical direction.)}

\end{center}
\end{figure}

Equivalent widths were measured independently by each of J.E.N. and
D.Y. for a set of unblended lines taken from the list of
\citet{cayrel04} (as described by \citet{norris10a} in our analysis of
the extremely metal-poor red giant {\boos}) with techniques described
by \citet{norris01} and \citet{yong08}. J.E.N. and D.Y. compared their
results and excluded from further consideration lines for which their
measured equivalent widths differed by more than 15 m{\AA}.  The
resulting two sets of line strengths are compared for the seven {\boo}
giants in the left panels of Figure~\ref{Fig:ew}.  While small
departures from the one-to-one line are evident in the figure,
representing a systematic difference of a few m{\AA}, NY chose to
simply average their two measursements. The resulting line strengths
in Table~\ref{Tab:nyeweps} for 115 unblended lines are suitable for
model atmosphere abundance analysis. Line identifications, lower
excitation potentials, $\chi$, and {\loggf} values are presented in
columns (1) -- (4) of the table.

\input{tab4e}

\subsubsection{GM measurements}\label{Sec:equivgm}

After sky-subtraction, the individual stellar spectra had radial
velocities determined as described above. For each star, the
individual spectra were then reduced to the rest-frame,
continuum-normalized and median-combined. In order to increase the
$S/N$ ratio used for the abundance analysis, the spectra were then
double-binned, obtaining a step of 0.028\,{\AA}\, and 0.034\,{\AA}\,
per pixel in the 4800 -- 5750\,{\AA} and 5840 -- 6800\,{\AA}\,
regions, respectively.  The spectra were further smoothed with a
Gaussian having standard deviation of 0.035\,{\AA}, which caused a
negligible loss of resolution, from the nominal initial R = 47000 to R
= 45000.

Equivalent widths were determined for a set of lines assembled from a
number of literature references (see \citealt{monaco11}), and for
which atomic parameters were obtained from the Vienna Atomic Line
Database
(VALD)\footnote{http://www.astro.uu.se/$\sim$vald/}\citep{kupka00},
except for Fe II, for which GM adopted the {\loggf} values of
\citet{melendez09}.  With one exception, the line strengths were
measured by using Gaussian fitting with the `fitline' code developed
by P.~Fran\c cois (private communication; see \citealt{lemasle07}).
All lines were inspected by eye and the continuum re-defined
interactively.  The code allows for deblending as well.  The exception
noted above was Na, for which measurement was achieved using
IRAF/splot.  For the Na lines, which have strengths greater than
150\,m{\AA}, Voigt rather than Gaussian profiles were fitted.  The
comparison between the line strengths of GM with those of NY is
presented in the right hand panels of Figure~\ref{Fig:ew}.
Table~\ref{Tab:gmeweps} contains the results from the GM analysis for
some 226 lines, where the format is the same as in
Table~\ref{Tab:nyeweps}.

\input{tab5e}

We note that both NY and GM discarded from analysis the (measured)
equivalent widths of the important [O~I]~6300.3\,{\AA} line. During
the span of our observations the radial velocity of {\boo}, together
with the Earth's orbital motion, positioned this line in the vicinity
of the telluric feature at 6302.0\,{\AA}, precluding reliable
determination of the stellar oxygen abundances.

\section{CHEMICAL ABUNDANCE ANALYSIS}\label{Sec:analysis}

Following the independent measurement of line strengths in
Section~\ref{Sec:equivalentwidths}, NY and GM performed abundance
analyses of their respective equivalent-width data sets.

\subsection{NY Analysis}\label{Sec:nyanalysis}

The NY model atmosphere analysis was as described by
\citet[Section~3]{norris10a}, to which we refer the reader for
details.  Suffice it here to say that NY adopted the ATLAS9 models of
\citet{castelli03}\footnote{http://wwwuser.oat.ts.astro.it/castelli/grids.html}
(plane-parallel, one-dimensional (1D), local thermodynamic equilibrium
(LTE)), with $\alpha$-enhancement, [$\alpha$/Fe] = +0.4, and
microturbulent velocity {\vxi} = 2~{\kms}.  These were used in
conjunction with the LTE stellar-line-analysis program MOOG
\citep{sneden73}; the version NY used includes an updated treatment of
continuum scattering (see \citealt{sobeck11}).  (We refer the reader
to \citet{cayrel04} and \citet{sobeck11} for discussions regarding the
importance of Raleigh scattering at blue wavelengths in metal-poor
stars.)  The only free parameter in the NY analysis is the
microturbulent velocity, {\vxi}, which was constrained by the
requirement that the abundance determined from the Fe I lines be
independent of their equivalent widths. In this process, NY deleted
from further analysis Fe I lines that fell more than either 3$\sigma$
or 0.5~dex from the mean value.

In \citet{norris10a}, NY tested their techniques by determining
abundances for the metal-poor stars observed and analyzed by
\citet{cayrel04}. They adopted as input the Cayrel equivalent widths,
atomic line data, and model atmosphere parameters, {\teff} and
{\logg}.  NY concluded that the agreement between their results and
those of \citet{cayrel04} was very satisfactory: the mean absolute
difference between relative abundances [X/Fe] for 11 elemental species
was 0.025 dex.  That is, the NY combination of MOOG +
\citet{castelli03} models produces essentially the same results as
Turbospectrum \citep{alvarez98} + OSMARCS \citep{gustafsson75} models
as utilized by \citet{cayrel04}.

\begin{figure*}[htbp]
\begin{center}
\includegraphics[width=12.0cm,angle=0]{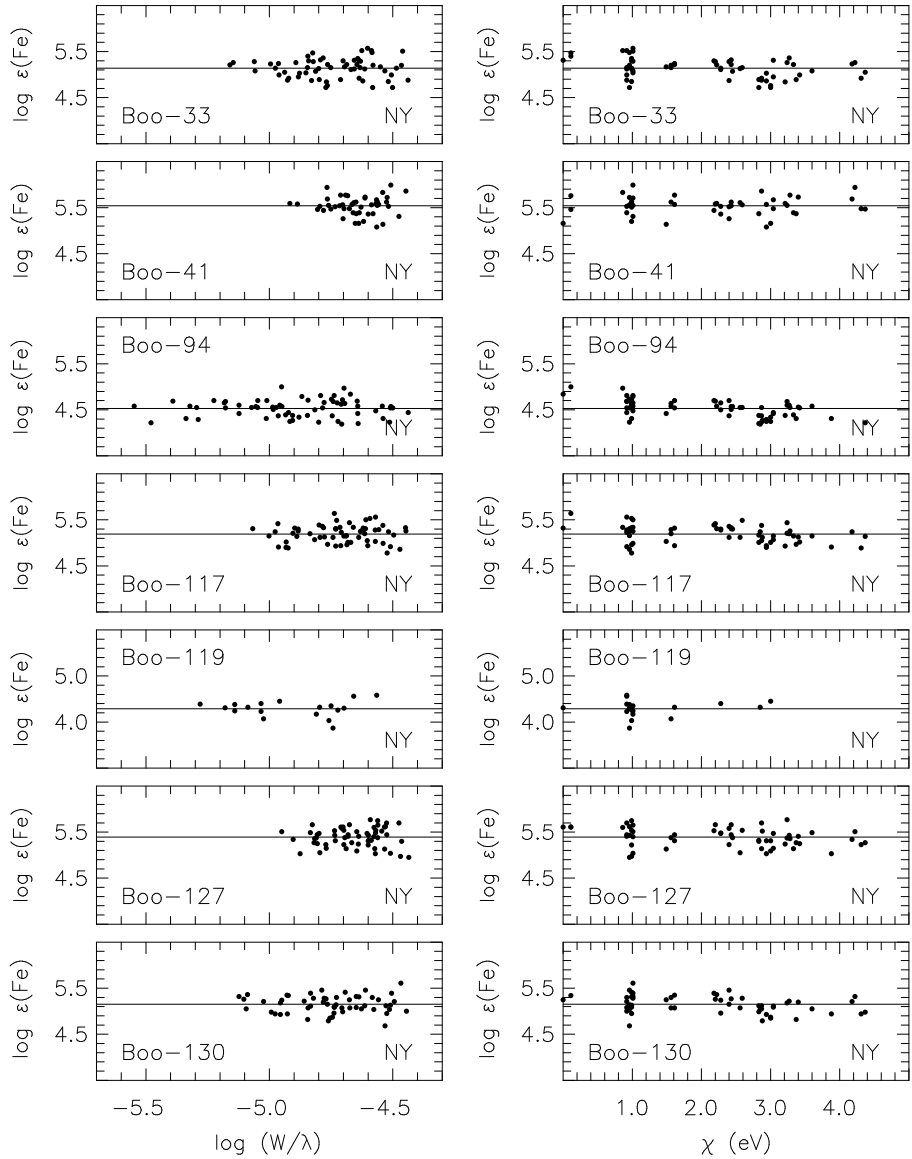}

  \caption{\label{Fig:nyeps}The abundances log\,$\epsilon$ of NY as a
    function of (left) log(W/$\lambda$) and (right) lower excitation
    potential ($\chi$) for the {\boo} giants.}

\end{center}
\end{figure*}

The NY absolute abundances, log\,$\epsilon$(X) ( = log(N$_{\rm
  X}$/N$_{\rm H}$) + 12.0), for individual lines in the seven {\boo}
red giants are presented in Table~\ref{Tab:nyeweps}, while [Fe/H] and
{\vxi} values are presented in column (4) and (5) of
Table~\ref{Tab:param}.  Figure~\ref{Fig:nyeps} shows these abundances
as a function of log(W/$\lambda$) (where W is the equivalent width)
and lower excitation potential, $\chi$.  Note the absence of any
dependence of abundance on either log(W/$\lambda$) or $\chi$.

\subsection{GM Analysis}\label{Sec:gmanalysis}    

The stellar atmospheric parameters {\teff} and {\logg}, the
determination of which is described above, were used by GM to
construct model atmospheres, using the Linux port of Version 9 of the
ATLAS code (\citealp{kurucz93a, kurucz93b, sbordone04}).

\begin{figure*}[htbp]
\begin{center}
\includegraphics[width=12.0cm,angle=0]{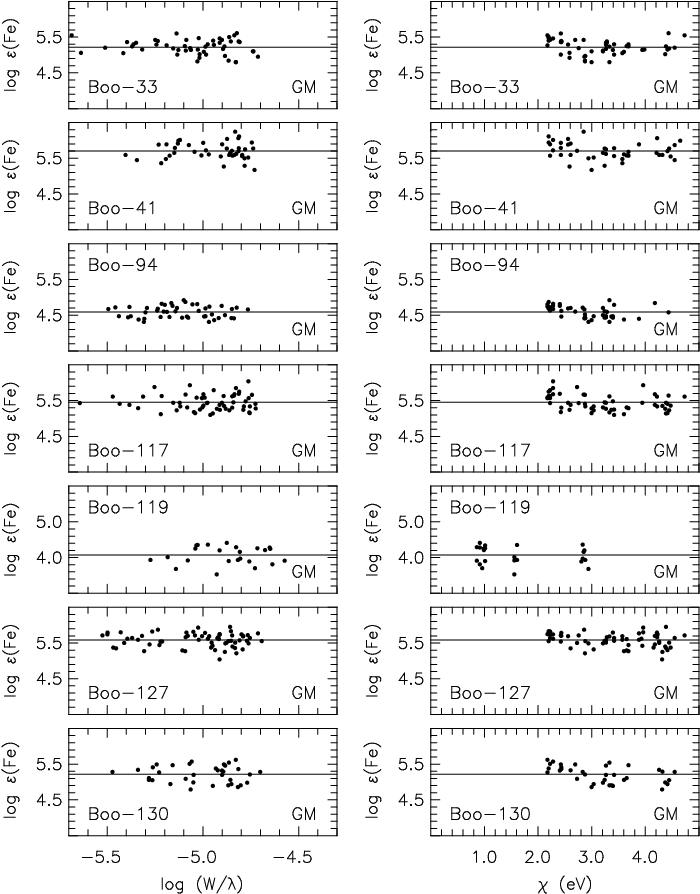}

  \caption{\label{Fig:gmeps}The abundances log\,$\epsilon$ of GM as a
    function of (left) log(W/$\lambda$) and (right) lower excitation
    potential ($\chi$) for the {\boo} giants.}

\end{center}
\end{figure*}

Chemical abundances were calculated using the Kurucz WIDTH9 code
together with the computed ATLAS9 model atmospheres and the measured
line strengths presented in Table~\ref{Tab:gmeweps}.  Microturbulent
velocities {\vxi} were adopted to minimize the dependence of the
abundance derived from Fe~I lines on their equivalent widths.  For all
stars but Boo-119, GM used only lines having EWs $<$ 100\,m{\AA} and
excitation potential $\chi$ $>2.0$\,eV, but given the very limited
number of lines detected in Boo-119, GM relaxed these two constraints
in that case.  The values of [Fe/H] and {\vxi} adopted by GM are
presented in columns (8) and (9) of Table~\ref{Tab:param}.  Note that
the GM microturbulent velocity for Boo-119, {\vxi} = 2.9 {\kms}, is
significantly higher than for the other six stars, {\vxi} $\sim$ 2.0
{\kms}, even though all stars have comparable values of gravity.
Adopting ${\xi}$ = 2.0 {\kms} for Boo-119 would have resulted in an
iron abundance that was 0.3 dex higher. The GM absolute abundance
obtained for each line in our {\boo} stars are presented in
Table~\ref{Tab:gmeweps}.  Figure~\ref{Fig:gmeps} shows abundances as a
function of log(W/$\lambda$) and lower excitation potential, $\chi$,
where no dependence on either log(W/$\lambda$) or $\chi$ in seen.

\subsection{Comparison of NY and GM Abundances}\label{Sec:comparison} 

Columns (3) -- (5) and (8) -- (10) of Table~\ref{Tab:xfe} contain mean
absolute abundances (log\,$\epsilon$), standard error of the mean
(s.e.$_{\log\epsilon}$), and number of lines analyzed, for 14 atomic
species from the NY and GM analyses, respectively.  In what follows we
shall be interested principally in the corresponding relative
abundances, [X/Fe] (for iron we tabulate [Fe/H]), which we present in
columns (6; NY results) and (11; GM results).  In order to determine
these values we have adopted the solar photospheric abundances of
\citet{asplund09}, which we also include for completeness in column
(2) of the table\footnote{A critic has suggested that we should draw
  the reader's attention to the fact that the NY [Fe/H] values in
  Table~\ref{Tab:xfe} agree well with those in Table 3 of
  \citet{norris10b} when allowance is made for the fact that in the
  earlier work the slightly different solar abundances of
  \citet{asplund05} were adopted.}.

\input{tab6}

\begin{figure*}[htbp]
\begin{center}
\includegraphics[width=15.0cm,angle=0]{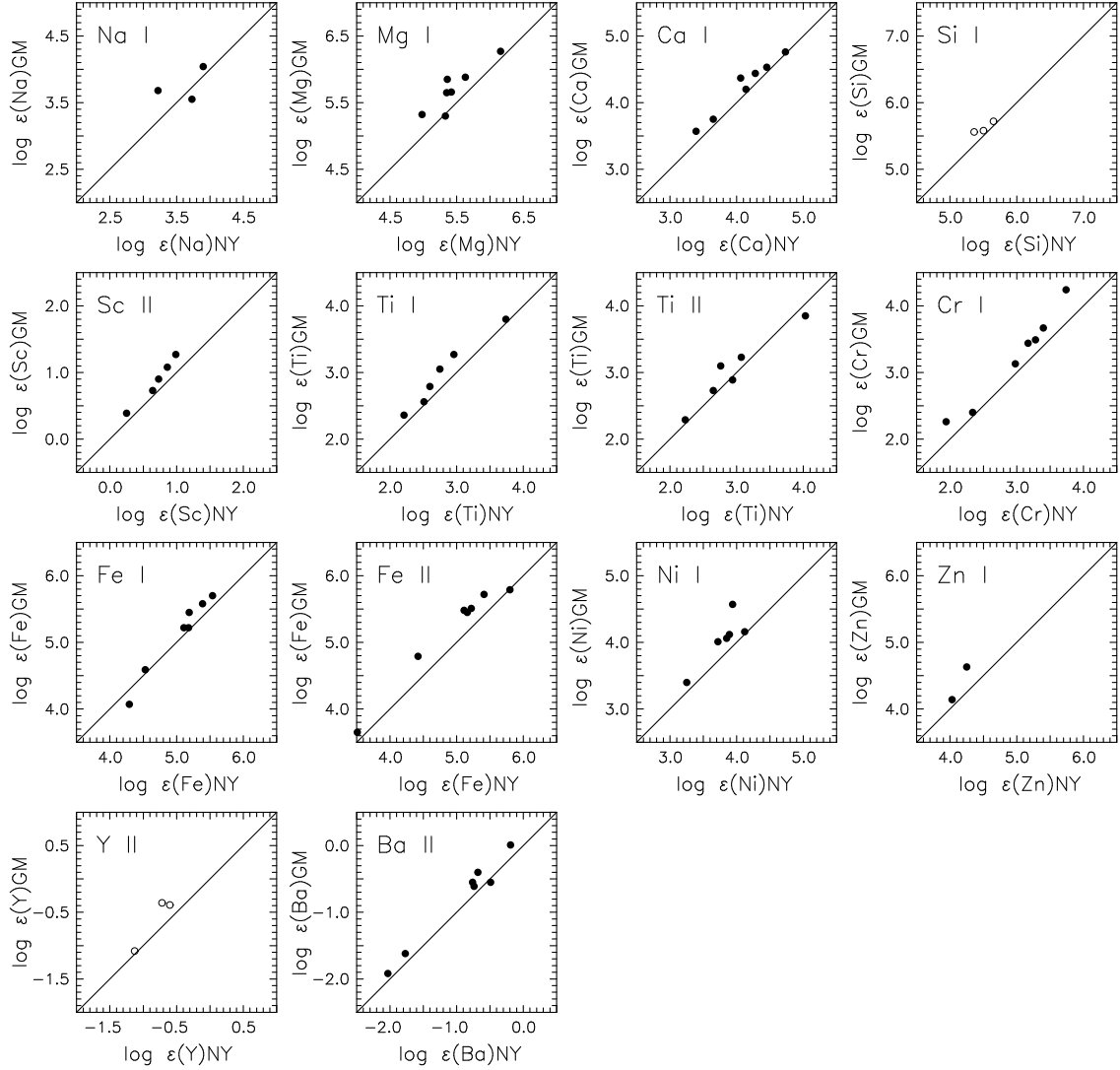}

  \caption{\label{Fig:logeps} The abundances log\,$\epsilon$ of GM
    vs. those of NY for 14 atomic species in the {\boo} giants. Filled
    circles refer to stars for which GM and NY both obtained
    abundances, while open circles (for Si~I and Y~II) represent those
    for which GM determined abundances while NY obtained only limits.}

\end{center}
\end{figure*}

A comparison of the log\,$\epsilon$ values from the two investigations
is presented in Figure~\ref{Fig:logeps}, where filled circles refer to
stars for which GM and NY both obtained abundances, while open circles
(for Si~I and Y~II) represent those for which GM determined abundances
while NY obtained only limits. NY chose not to measure equivalent
widths for these elements given the weakness of the lines and the
quality of the spectra. {\it Post facto}, NY measured line strength
limits for the Si\,I~5948.54\,{\AA} and Y\,II~4883.68\,{\AA} lines in
those stars for which GM reported detections.  The limits shown in
Figure~\ref{Fig:logeps} for Si\,I and Y\,II were obtained by using the
$gf$ values adopted by GM.

After the independent NY and GM analyses described above had been
completed, we sought to understand the sources of the (mostly small)
abundance discrepancies between them.  We first tested for differences
that might result from the adopted combinations of model atmospheres
and emergent flux code: NY use \citet{castelli03} models + MOOG
(Section~\ref{Sec:nyanalysis}), while GM adopt ATLAS9 models computed
by them + WIDTH9 (Section~\ref{Sec:gmanalysis}).  We found that when
each group adopted the input data ({\teff}, {\logg}, microturbulence,
atomic parameters, {\loggf} values, and equivalent widths) of the
other, it reproduced the abundances of the other extremely well.  For
example, when NY analyzed the data of GM for the Fe\,I lines they
obtained mean differences
$\langle$$\Delta$log\,$\epsilon$$\rangle$(NY--GM) in the range
$-0.029$ to +0.035 dex for the seven stars, with dispersions in the
range 0.007 to 0.020 dex.

We examined the differences that might be driven by the choice of $gf$
values. Although in general agreement is good, when only a small
number of lines is available for an atomic species, real differences
occur, which can be explained in terms of a different choice of $gf$
values. An example of this is seen in the Mg panel in
Figure~\ref{Fig:logeps} (and in Figure~\ref{Fig:feltzing} in the
following section) where one sees that NY determine lower abundances
than GM, driven in large part by NY and GM adopting {\loggf} = --0.34
and --0.62 for Mg~I 5528.40~{\AA}, respectively.  In other cases, as
expected, differences in measured equivalent width resulting from
spectrum signal-to-noise limits, in particular for weak lines, are
responsible for the discrepancies.

Inspection of the log\,$\epsilon$, [Fe/H], and {\vxi} values in
Tables~\ref{Tab:param}, \ref{Tab:nyeweps}, and \ref{Tab:gmeweps} shows
that while the average of the iron abundance difference between the
results on NY and GM for the seven {\boo} members is $-0.09$ $\pm$
0.06 dex, the corresponding average of microturbulent velocity
difference ($\Delta${\vxi}, in the sense NY -- GM) is +0.6 $\pm$ 0.2
{\kms}.  Insofar as one may infer from Section~\ref{Sec:errors}
(Table~\ref{Tab:errors}) below that a change of +0.6 {\kms} will cause
a change $\Delta$[Fe/H]~=~--0.2~dex, this value of the average
difference is somewhat larger than expected.  We believe the effect
can be understood, in large part, by the fact that while NY analyze
all lines with equivalent widths less than 200~m{\AA}, GM exclude
those stronger than 100 m{\AA} and with excitation potential less than
2~eV.  When we reanalyze the NY data set using the above GM limits on
line strength and excitation potential we find that while the mean
abundance difference remains small at --0.01 $\pm$ 0.08~dex, the
difference in microturbulence decreases to
$\langle\Delta${\vxi}$\rangle$ = 0.2 $\pm$ 0.3 {\kms}.  That is to
say, the derived value of microturbulence appears to depend on the
upper line strength and the lower excitation potential of the sample
of lines that we have chosen to analyze, while, on the other hand,
[Fe/H] remains essentially unchanged.

\input{tab7}

Two effects that might contribute to the difference are as follows.
First, microturbulence is an artifact introduced to explain the
deficiency of one-dimensional model atmosphere analyses: it is not
needed, for example, in three-dimensional models of the Sun
\citep{asplund00}.  One might not be surprised to find that lines of
different strength, well away from the linear part of the
curve-of-growth, might need different amounts of correction.  Second,
one should also consider the possibility that since NY have employed
Gaussian fitting in their measurement of equivalent width, they might
have underestimated the strengths of the lines in the range 100 -- 200
m{\AA}.  Against this possibility, we note that for the three Na I
lines that have equivalent widths greater than 150 m{\AA} in both
Tables~\ref{Tab:nyeweps} and \ref{Tab:gmeweps} and for which NY and GY
fit Gaussian and Voigt profiles, respectively, the mean line strength
difference is only 3 m{\AA}.  That said, insofar as the measured
difference in microturbulence have no effect on the conclusions we
reach about the abundances in our sample, we shall not explore these
possibilities further. 

We conclude that where there were systematic differences in the
adopted methods and/or assumptions, systematic differences between the
authors' abundances follow. These systematic effects can by nullified
by suitable adoption of an internally consistent analysis approach to
a single data set. Random errors which remain are driven by spectrum
signal-to-noise limitations. However, the systematic, analysis
methodology-dependent differences detected here make clear that data
sets from different authors cannot simply be combined to detect -- or
limit -- intrinsic abundance dispersions in stellar element abundances.

\subsection{Internal, External and Adopted Abundance Uncertainties}\label{Sec:errors}

The appropriate random internal error of the absolute abundances in
columns (3) and (8) of Table~\ref{Tab:xfe} is the standard error of
the mean of the several lines analyzed per element by each analysis
team, s.e.$_{\log\epsilon}$, which we have presented in columns (4)
and (9) of the table, respectively.

These abundances are also potentially subject to systematic
uncertainties resulting from the uncertain atmospheric
parameters. This uncertainty cannot be determined accurately, but may
be approximated as follows.  Starting with a model atmosphere having
{\teff} = 4750\,K, {\logg} = 1.4, [Fe/H] = --2.0, and {\vxi} =
1.8~{\kms} we varied the relevant parameters, one at a time, by
$\Delta${\teff} = $\pm$100\,K, $\Delta${\logg} = $\pm$0.3, and
$\Delta\xi_t$ = $\pm$0.3~{\kms}.  These parameter changes correspond
to twice the amplitude appropriate for {\teff} and for {\logg}, and
the amplitude appropriate for {\vxi} deduced from the discussion in
the previous section, based on the parameter ranges determined by NY
and by GM for individual stars in Table~\ref{Tab:param}. That is,
these external error estimates are conservative (and assume that
covariance errors terms are negligible).  Since we shall be interested
mainly in relative abundances [X/Fe], we have determined the
corresponding uncertainties in [X/Fe] (for iron we estimated the
uncertainty in [Fe/H]).  Our listed estimates of the elemental
abundance uncertainties associated with uncertainties in stellar
atmospheric parameter determination are presented in
Table~\ref{Tab:errors}. Columns (2) -- (4) contain the scalings of the
individual contributions to the errors, and the final column shows the
accumulated uncertainty when the three errors are added
quadratically. We note this is again a conservative approach, as many
of the systematic potential errors have opposite sign, and can
cancel. Quadratic addition does not allow for this.

To obtain total error estimates, we adopted the following procedure
(cf. \citealt{norris10a}).  The random errors in columns (4) and (9)
of Table~\ref{Tab:xfe} are internal formal estimates of the underlying
appropriate error distribution, based on the dispersion in what is
often a small number of lines, and hence is itself uncertain. We
replace this estimated random error, s.e.$_{\log\epsilon}$, from $N$
lines, by max(s.e.$_{\log\epsilon}$, s.e.$_{\log\epsilon(\rm Fe\,I)}$
${\times}$ $\sqrt{N_{\rm Fe I}/N}$).  The second term is what one
might expect from a set of $N$ lines having the dispersion we obtained
from our more numerous ($N_{\rm Fe I}$) Fe I lines.  We then
quadratically combined this updated random error and the error
associated with uncertainty in the atmospheric parameters from column
(5) in Table~\ref{Tab:errors} to obtain the total error,
$\sigma$[X/Fe], which we present in columns (7) and (12) of
Table~\ref{Tab:xfe}, for the NY and GM analyses, respectively.
 
\begin{figure}[htbp]
\begin{center}
\includegraphics[width=7.0cm,angle=0]{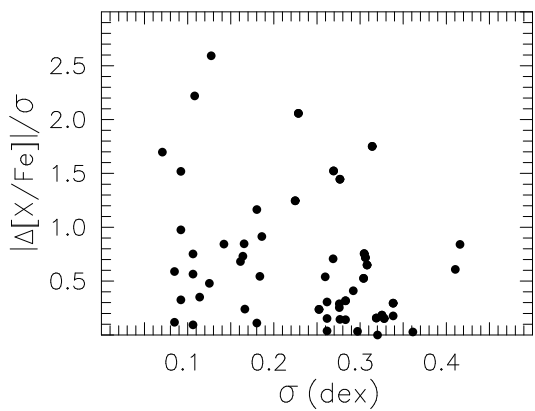}

  \caption{\label{Fig:delsigs} $|\Delta$[X/Fe]$|$/$\sigma$ as a
    function $\sigma$, where $\Delta$[X/Fe] = [X/Fe]$_{\rm NY}$ --
    [X/Fe]$_{\rm GM}$ and $\sigma$ = ($\sigma_{\rm NY}^{2}$ +
    $\sigma_{\rm GM}^{2})^{0.5}$, are the measured and expected differences
    between the analyses of NY and GM.  See text for discussion.}

\end{center}
\end{figure}

Consideration of Figure~\ref{Fig:logeps} and Table~\ref{Tab:xfe} shows
the agreement between the NY and GM analyses is in general consistent
within the quoted uncertainties. Indeed, for stars having relative
abundances determined for the same species by both NY and GM, the mean
difference of $\Delta$[X/Fe] = [X/Fe]$_{\rm NY}$ -- [X/Fe]$_{\rm GM}$
is --0.09~dex, while the median absolute difference is 0.09~dex.  It
is also instructive to compare the abundance difference $\Delta$[X/Fe]
with the error $\sigma$ = ($\sigma_{\rm NY}^{2}$ + $\sigma_{\rm
  GM}^{2})^{0.5}$ expected from the total error estimates of NY and GM
in columns (7) and (12) of Table~\ref{Tab:xfe}.  This is done in
Figure~\ref{Fig:delsigs} which presents $|\Delta$[X/Fe]$|$/$\sigma$
vs. $\sigma$.  Here some 80\% and 95\% of points fall below
$|\Delta$[X/Fe]$|$/$\sigma$ = 1 and 2, respectively, in excellent
agreement with expectation.

\subsection{Comparison with the Abundances of \citet{feltzing09}}
\label{Sec:feltzing}  

\begin{figure*}[htbp]
\begin{center}
\includegraphics[width=17.0cm,angle=0]{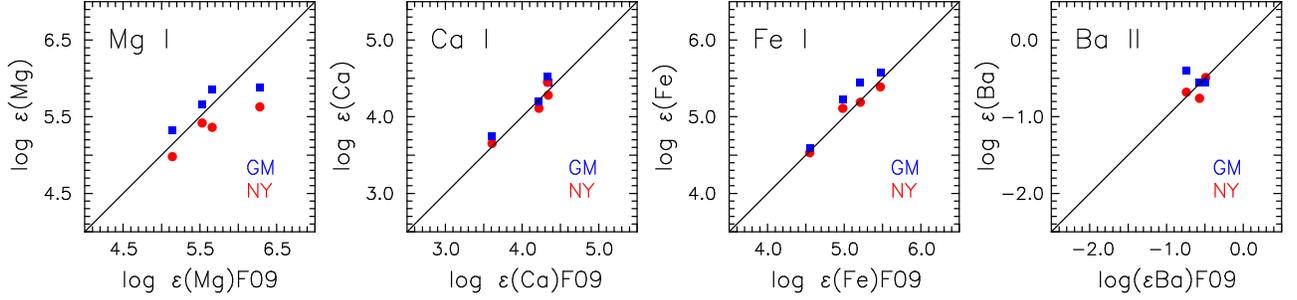}

  \caption{\label{Fig:feltzing} The abundances log\,$\epsilon$ from
    the present work of NY (red circles) and GM (blue squares)
    vs. those of \citet{feltzing09} for elements in common between the
    two investigations.}

\end{center}
\end{figure*}

Four of the stars observed here have also been analyzed by
\citet{feltzing09}, who presented abundances of Mg, Ca, Fe, and Ba for
them.  These stars are identified in Table~\ref{Tab:param} where we
include the \citet{feltzing09} values of {\teff}, {\logg}, [Fe/H], and
{\vxi} in columns (10) -- (13) (there labeled F09).  One sees that the
stellar astrophysical parameters are in agreement within those of the
present work, given the rather coarse parameter grid adopted by
\citet{feltzing09}.  Figure~\ref{Fig:feltzing} compares our
log\,$\epsilon$ values with theirs, where red circles and blue squares
refer to the results of NY and GM, respectively.  For Ca, Fe, and Ba
the abundance agreement is excellent.  For Mg, however, there is one
discrepant object, Boo-127, for which \citet{feltzing09} report
[Mg/Fe] $\sim$ +0.7 and [Mg/Ca] $\sim$ +0.7~dex.  We shall discuss
this discrepancy further in Section~\ref{Sec:relabund}.

The conclusion of this external literature comparison is that the
internal accuracy of elemental abundances determinable from spectra of
the quality available here is adequately represented by the
uncertainties listed in Table~\ref{Tab:xfe}, while single discordant
measures should be treated as provisional.

\section{RELATIVE ELEMENT ABUNDANCES}\label{Sec:relabund}

\begin{figure*}[htbp]
\begin{center}
\includegraphics[width=14.0cm,angle=0]{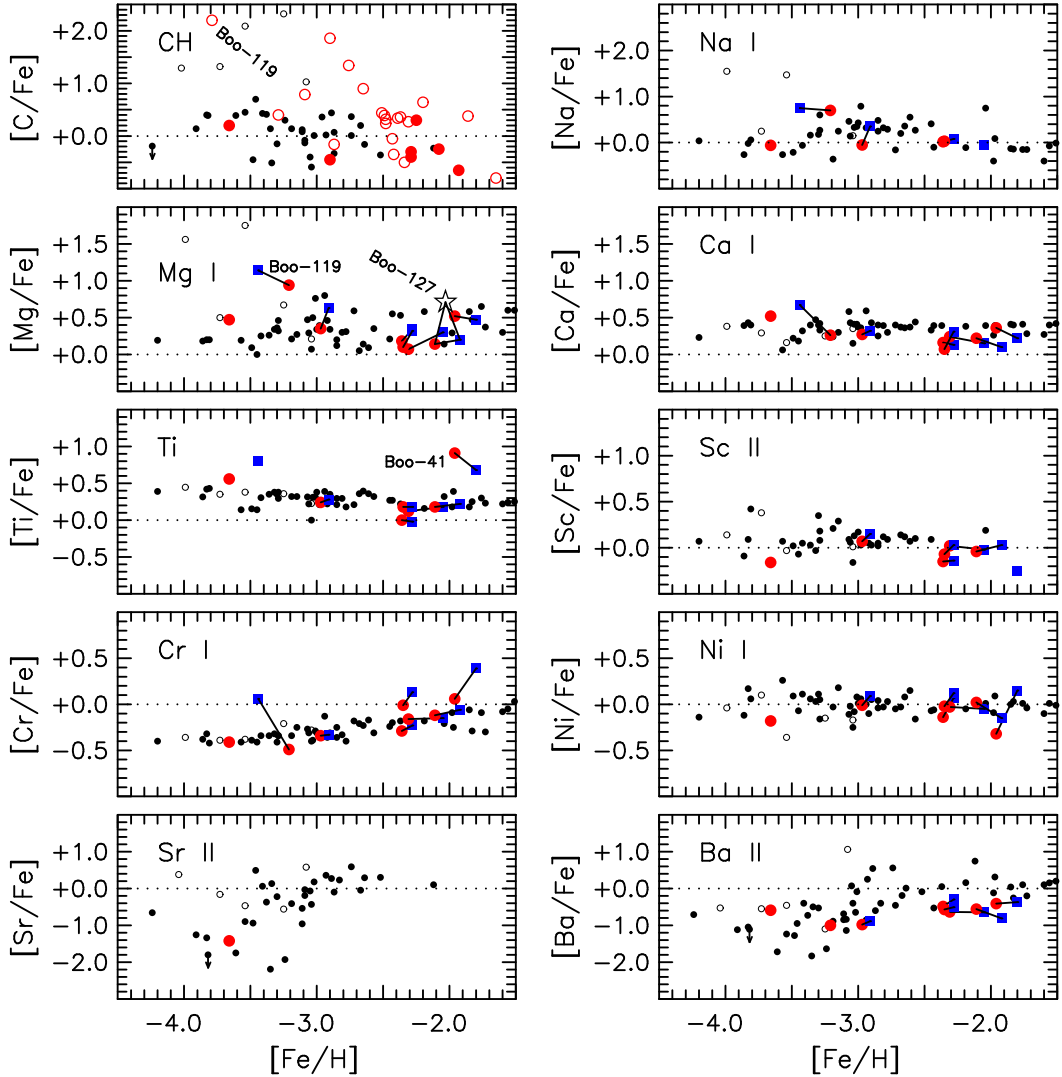}

  \caption{\label{Fig:xfe} \small Relative abundances [X/Fe]
    vs. [Fe/H] for {\boo} and Galactic halo red giants.  (The observed
    species is identified in each panel, and Ti represents the average
    of Ti\,I and Ti\,II.)  In all panels except that at top left, the
    large blue filled squares represent the abundances of GM, while
    the large red filled circles stand for those of NY and {\boos}
    (from \citealt{norris10a}) (Results for the same star are
    connected).  In the upper left panel the large filled and open red
    circles represent the carbon and iron abundances of
    \citet{norris10b} and \citet{lai11}, respectively, for {\boo} red
    giants.  In all panels the small filled and open black circles
    represent Galactic halo C-normal and C-rich stars, respectively,
    from the work of \citet{aoki02, aoki04}, \citet{cayrel04},
    \citet{depagne02}, \citet{francois07},
    \citet{fulbright00}, \citet{ito09}, \citet{norris97}, and
    \citet{spite05}.  In the Mg I panel, the black star presents
    results for Boo-127 from the work of \citet{feltzing09}; see
    Sections~\ref{Sec:feltzing} and \ref{Sec:mgca} for discussion of
    this object.}

\end{center}
\end{figure*}

The dependence of relative abundances, [X/Fe], on [Fe/H] for {\boo} is
shown in Figure~\ref{Fig:xfe} as large red filled circles (NY)
connected to large blue filled squares (GM), together with the data
for {\boos} from \citet{norris10a} (as a large red filled circle at
[Fe/H] = --3.7).  In the top left panel we plot the {\boo} [C/Fe] and
[Fe/H] abundances from \citet{norris10b} (large filled red circles)
and \citet{lai11} (large open red circles).  For comparison, we also
present the abundances for C-normal and C-rich red giants of the
Galactic halo (as small filled and open black circles, respectively)
from the work of \citet{aoki02, aoki04}, \citet{cayrel04},
\citet{depagne02}, \citet{francois07}, \citet{fulbright00},
\citet{ito09}, \citet{norris97}, and \citet{spite05}\footnote{For Na
  -- Ba, abundances of which have been given in our
  Table~\ref{Tab:xfe}, we have modified the literature values
  presented in Figure~\ref{Fig:xfe} (and in Figure~\ref{Fig:alpha8}
  below) to take into account the different solar abundances adopted
  in those works, in order to place them on the \cite{asplund09} scale
  used here.  We have not, however, attempted to do this for carbon,
  where we assume the differences will be small relative to the large
  [C/Fe] range of 3.5~dex in Figure~\ref{Fig:xfe}.}.  Inspection of
  Figure~\ref{Fig:xfe} confirms that, with the exception of the
  abundances for Boo-119 (at [Fe/H] $\sim$ --3.3) and for Cr I in two
  other stars, the results of the NY and GM abundance determinations
  are in agreement, and the uncertainties derived above are a fair
  estimate of the true error.  The figure also gives one the general
  impression that to first order the abundances of the {\boo} giants
  follow the trends established by the Galactic halo giants, as we
  reported earlier for {\boos} \citep{norris10a}.

Comparison with the halo trends identifies two anomalies in the data
presented in Figure~\ref{Fig:xfe}.  The first is the large departure
of [Ti/Fe] in Boo-41 (at [Fe/H] $\sim$ --1.85) from the
trends found in both {\boo} and the halo of the Milky Way.  In
contradistinction, the [Ti/Fe] values of NY and GM agree well, for
both Ti\,~I and Ti\,~II.  We can offer no explanation for the
discrepancy, which is particularly puzzling, since, as we shall see
below, Boo-41 is not an outlier when one considers [Ca/Fe], which has
the smallest errors by far of the three $\alpha$-elements (Mg, Ca, Ti)
observed here.

The second significant discrepancy is between the NY and GM results
for Boo-119. For the four species investigated in both analyses -- Na,
Mg, Ca, and Cr -- the absolute differences in $\Delta$[X/Fe] are 0.05,
0.20, 0.41, and 0.54~dex, respectively.  We refer the reader to
Section~\ref{Sec:comparison} for the discussion of effects that might
lead to differences such as these. Given our conclusion that Boo-119
is a member of the rare CEMP-no class, further observations should be
obtained to improve and extend the present results.

\subsection{[Mg/Fe] and [Mg/Ca] Enhancements in {\boo}?}\label{Sec:mgca}

There have been a number of reports of significantly non-solar [Mg/Fe]
and [Mg/Ca] values in otherwise normal stars in the metal-poor
populations of the Galaxy and its satellite galaxies, ranging from
[Mg/Ca] = --1.2 in a halo field red giant \citep{lai09} to +0.9 in the
Hercules {\uf} dwarf galaxy \citep{koch08}.  As noted in
Section~\ref{Sec:feltzing}, \citet{feltzing09} reported [Mg/Fe] $\sim$
+0.7 and [Mg/Ca] $\sim$ +0.7 for Boo-127.  We are unable to confirm
these results for Boo-127 in the present work.  The discrepancy is
shown in the [Mg/Fe] panel of Figure~\ref{Fig:xfe} where we represent
their result as an open star, joined by lines to the present values of
NY and GM, both of whom do not find it to have enhanced [Mg/Fe]
compared with the Galactic halo trend. We note that we also find
[Mg/Ca] = --0.08 (NY) and +0.09 (GM), essentially the solar value.
Given the importance of the interpretation of variations in [Mg/Ca] as
the signature of enrichment by individual supernovae, confirming an
anomalous value requires strong evidence. Here we advocate that our
determination of [Mg/Ca] in this star be adopted, unless/until future
new data support an anomalous value.

\begin{figure}[htbp]
\begin{center}
\includegraphics[width=8.5cm,angle=0]{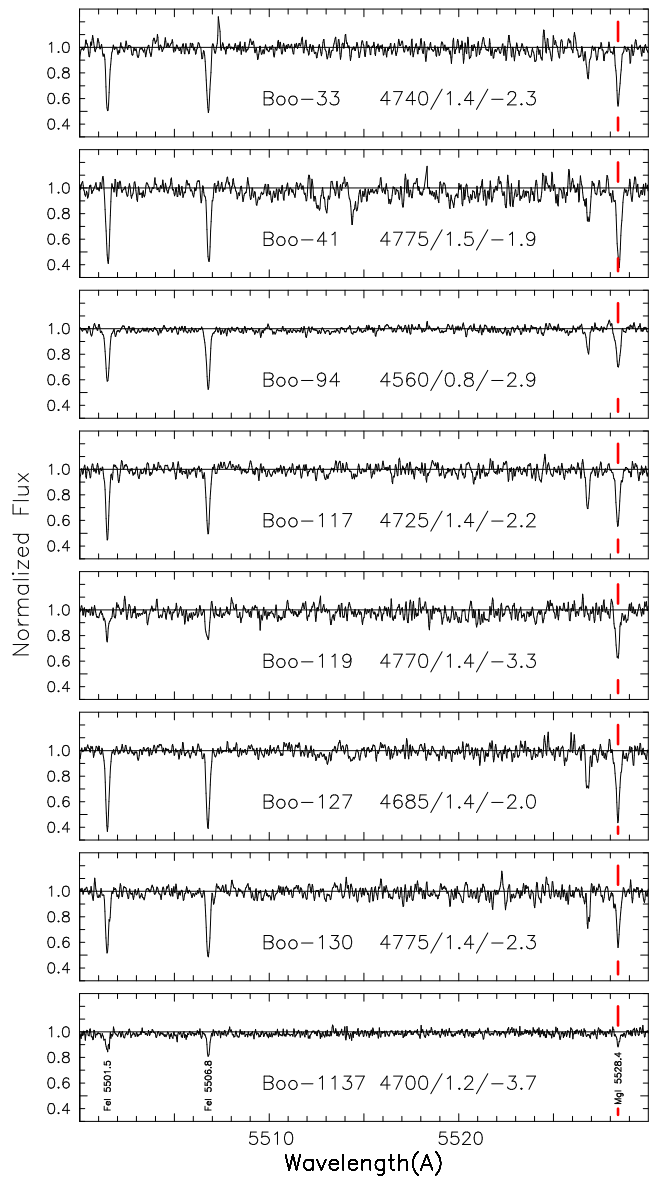}

  \caption{\label{Fig:mgii5528} Spectra of the {\boo} giants in the
    region of Mg I 5528.4\,{\AA}.  Three atomic lines are identified
    in the lowest panel. Note that only in Boo-119 is the Mg I line
    stronger than the two lines Fe I 5501.5 and 5506.8\,{\AA}.
    Boo-119 is Mg-rich, with [Mg/Fe] = 1.0.  Mean adopted atmospheric
    parameters {\teff}/{\logg}/[Fe/H] are shown in each panel.}

\end{center}
\end{figure}

In the [Mg/Fe] panel of Figure~\ref{Fig:xfe}, we identify the outlier
Boo-119, with [Mg/Fe] $\sim+1.0$ ([Mg/Ca] = +0.7).  To support the
reality of this measurement we present our spectra of the {\boo}
giants in the region of the Mg~I 5528.4\,{\AA} line in
Figure~\ref{Fig:mgii5528}, where in the lowest panel the Mg~I line and
two Fe~I lines are identified.  Inspection of the figure shows that
the Mg~I line is significantly stronger than the Fe I lines in the
spectrum of Boo-119, and in only this star.

Fortuitously, additional abundance information is available for
Boo-119: it has also been observed by \citet{lai11}, who designate it
Boo21, and report abundances [Fe/H] = --3.8 and [C/Fe] = +2.2 (this is
the C-rich star at top left of the [C/Fe] panel in
Figure~\ref{Fig:xfe}; note that the low-resolution spectra obtained by
Lai et al.~do not provide abundances of individual $\alpha$-elements).
Boo-119 is thus very carbon rich.  It also has [Ba/Fe] = --1.0
(according to NY; GM do not measure this feature) and is therefore a
CEMP-no star (see \citealt{beers05} for the definition of terms; as
discussed below, the abundances measured in CEMP-no stars most likely
reflect those of the interstellar medium (ISM) from which they
formed). Mg enhancement is frequently observed in such stars
\citep{masseron10, norris12}.  We therefore conclude that the Mg
enhancement we report for Boo-119 is real, and that its C, Mg, and Ba
abundances are together consistent with its being a CEMP-no
star. Boo-119 joins Segue 1-7 as the second CEMP-no star to be
recognized in the Milky Way's {\uf} galaxies
(cf. \citealp{norris10c}).

\subsection{The $\alpha$-elements}\label{Sec:alphas}

\begin{figure}[htbp]
\begin{center}
\includegraphics[width=6.0cm,angle=0]{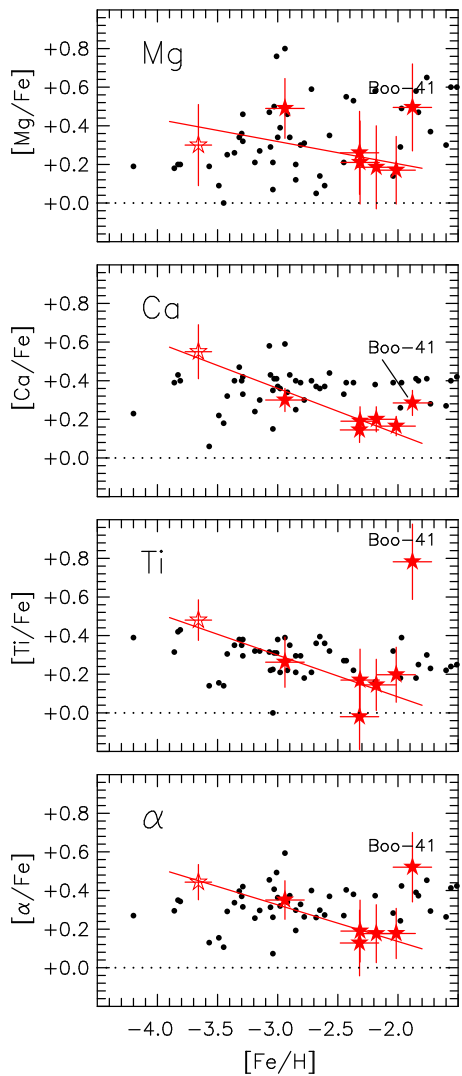}

  \caption{\label{Fig:alpha8} [$\alpha$/Fe] vs. [Fe/H] for C-normal
    red giants in {\boo} (filled red stars, this work; open red star,
    \citealp{norris10a}) and the Galactic halo (small black
    circles, from \citealp{cayrel04} and \citealp{fulbright00}). See
    text for discussion. }

\end{center}
\end{figure}

The relative abundances of the $\alpha$-elements [Mg/Fe], [Ca/Fe],
[Ti/Fe] (= ([Ti\,I/Fe] + [Ti\,II/Fe])/2), and [$\alpha$/Fe] (= [Mg/Fe]
+ [Ca/Fe] + [Ti/Fe])/3) are presented as a function of [Fe/H] in
Figure~\ref{Fig:alpha8}, for the stars of the present work (excluding
the Mg-enhanced CEMP-no star Boo-119), together with results for
{\boos} from \citet{norris10a}.  (For the stars in the present work,
we have averaged the abundances of NY and GM in Table~\ref{Tab:xfe},
while for {\boos} we have recomputed the $\alpha$-element abundances
using only lines from \citet{norris10a} having wavelength greater than
4870\,{\AA}, in order to reproduce more closely the wavelength
coverage of the present investigation.)
The data are summarised in Table~\ref{Tab:alphas}.

\input{tab8}

The relative abundances of the $\alpha$-elements to iron in the seven
carbon-normal (i.e., excluding Boo-119) member stars of {\boo} are, with
the exception of one outlier in one element, indistinguishable from
those of a typical star in the halo of the Milky Way, being $\lesssim
2\sigma$ away from the bulk of the field halo.  The outlier is Boo-41,
which, as noted above, is significantly more enhanced in Ti than is the
bulk of the halo, and indeed is more enhanced than the other member
stars of {\boo}.  We note for completeness that the four {\boo} stars
with [Fe/H] $\sim -2.25$, again excluding Boo-41 from the comparison,
have remarkably similar elemental abundances.

We note explicitly one pattern in the $\alpha$-element data which,
while of limited statistical significance, is a hint of what one would
search for in larger samples of {\boo} stars.  If one were to exclude
all the abundance data for Boo-41, on the basis of its anomalous Ti
abundance, and those data for Boo-119, the CEMP-no star, one sees a
clear decrease in the relative abundance of all three elements in the
remaining 6 member stars as [Fe/H] increases.  The line in each panel
of Figure~\ref{Fig:alpha8} is the least-squares line of best fit,
excluding Boo-41. For Mg, Ca, Ti, and $\alpha$ the slopes of the line
are --0.115, --0.237, --0.216, --0.189, respectively, while the RMS
scatters about the line are 0.105, 0.046, 0.105, and 0.043 dex.

Dependencies of this type are physically plausible, are suggestive of
a relatively low star formation rate at the earliest times in this
{\uf} system, and are driven by a contribution of Type~Ia supernovae
to the chemical enrichment of these elements. Such patterns are seen
in the more luminous dwarf spheroidal galaxies, albeit at higher
values of [Fe/H], where the higher iron abundance reflects faster
enrichment in these larger systems (see, e.g., \citealp[their
  Figure~11]{tolstoy09}).

This digression aside, retention of Boo-41 in our full sample of seven
carbon-normal {\boo} stars provides a different picture, with no clear
trend in $\alpha$-abundance ratios.  Rather, there is a distribution
which is, within measurement errors, consistent with having a mean
value, and a scatter about that mean, which are similar to those found
in the field halo, albeit with the exception of a significant scatter
in Ti\footnote{A referee has suggested that the anomalous behavior of
  some elements, in particular Ti, might have its origin in our use of
  LTE rather than non-LTE analysis techniques.  Unfortunately, the
  only relevant contribution on non-LTE Ti abundances of which we are
  aware is that of \citet{bergemann11} who, in the context of her
  analysis of four metal-poor stars, states: ``The Ti non-LTE model
  does not perform ... well for the metal-poor stars ...  we find that
  only [Ti/Fe] ratios based on Ti II and Fe II lines can be safely
  used in studies of Galactic chemical evolution''.  In our view, the
  quality of our Fe II abundances is insufficient to address this
  issue.  Our position is that LTE analysis currently presents a
  useful approach in the sense that if, in the ([X/Fe], [Fe/H]) --
  plane, one see distinct differential behavior between two groups of
  stars (in this case the Bootes I stars and the Galactic halo field
  stars), the source of the difference more likely lies in the
  chemical abundances of the two groups rather than within the
  approximations made in the analysis of their spectra.}.  If one
decided to sub-select the sample of {\boo} stars, one might even argue
for the exclusion of Boo-1137, which is presently at significantly
larger projected distance from the center than are the other stars (it
is at 24 arcmin, the rest within 8 arcmin); this would weaken the case
for a smooth decline in elemental ratios as the metallicity
increases. Data for a larger sample of {\boo} stars is clearly
desirable. That said, the lack of significant scatter in the entirety
of the present sample (further quantified in the following section) is
itself consistent with a relatively slow early star-formation (and
enrichment) rate, since time is required for SNe ejecta from a range
of progenitor masses to be created and mixed into the ISM prior to the
bulk of star formation.

\subsection{Relative-Abundance Dispersions}\label{Sec:dispersions}

A critical consideration in seeking to understand the manner in which
chemical enrichment occurs in any stellar system is the dispersion of
abundance as a function of overall enrichment.  To address this issue,
we follow \citet[their Section~4]{cayrel04}, who, for element X,
determine the linear least squares fit of [X/Fe] as a function of
[Fe/H] and measure the dispersion $\sigma$[X/Fe] about that fit.

We restrict the sample for this analysis of scatter to only the six
carbon-normal stars analyzed double-blind in this study (thus
excluding Boo-119 and Boo-1137).  Given the small sample of {\boo}
stars and in some cases very few lines of some elements, we have
further restricted our focus towards those atomic species for which we
can determine relative abundances most accurately.  To this end, we
included only neutral species, to minimize errors of measurement in
[X/Fe], and used only elements for which more than one line has been
measured in more than two stars. These conditions are met by Ca\,I,
Ti\,I, Cr\,I and Ni\,I.

Table~\ref{Tab:dispersions} presents the resulting dispersions, where
for each species columns (2) -- (5) contain the abundance dispersion,
the mean error of measurement, the number of stars, and the mean
number of lines measured, from the analysis of NY, while columns (6)
-- (9) contain the results from the GM analysis.  For comparison, we
also determined the same parameters from the abundances of
\citet{cayrel04}, for stars having --3.0 $<$ [Fe/H] $<$ --2.0 (the
range pertaining to the {\boo} stars under discussion here) and
present the results in columns (10) -- (13) of the table.  We
interpret the equality of the observed dispersion and errors of
measurement in the \citet{cayrel04} data set as indicating that no
intrinsic spread has been observed for these elements in halo stars in
this abundance range.  We also note that the smaller measurement
errors of \citet{cayrel04} compared with those in the present work are
not unexpected, given the higher signal-to-noise of their work ($S/N$
$\sim200$ per 0.015\,{\AA} pixel at 5100\,{\AA}) compared with our
value of $\sim30$ per 0.03\,{\AA} pixel at 5500\,{\AA}.

\input{tab9}

The data in Table~\ref{Tab:dispersions} provide no evidence for
detection of a spread in elemental ratios at any value of [Fe/H] in
{\boo}, with the already noted exception of Ti. Given the differences
between the NY and GM analyses, which make clear our quoted values of
$\sigma_{\rm meas}$ are estimates, not precision determinations, the
observed dispersions are commensurate with the errors of measurement.
Quadratic subtraction of these two quantities to produce intrinsic
dispersions, at better than the 0.05 -- 0.10~dex level, is
questionable.  Bearing this caveat in mind, we offer the following
comments.  For Ca, the NY and GM analyses admit intrinsic dispersions
of 0.08 and 0.05~dex, respectively.  We also recall from the previous
section that the average of the NY and GM relative abundances
(excluding the outlier Boo-41) yield an observed dispersion in [Ca/Fe]
of 0.046 dex.  For Ti, the data indicate a dispersion $\sim$0.25 dex,
driven largely by the inclusion of the outlier Boo-41 discussed above;
exclusion of this object reduces the intrinsic dispersion to
$\sim$0.1~dex.  For Cr, any intrinsic dispersion is poorly determined,
and lies in the range 0.00 -- 0.20~dex.  Finally, for Ni the more
accurate determination of GM suggest a dispersion not larger than
0.1~dex.

For completeness, we note that for the ionized species, which were
excluded from consideration by the selection criteria above
(i.e. Sc\,II, Ti\,II, and Ba\,II), the mean error of measurement lies
in the range 0.19 -- 0.24~dex.

\subsection{Barium}

For the {\boo} giants, the dispersion of [Ba/Fe] in
Figure~\ref{Fig:xfe} is consistent with that of the Galactic halo.  It
is interesting to compare the {\boo} data with those of
\citet{frebel10} for the Com Ber and UMa~II {\uf} dwarf galaxies.  For
Com Ber at [Fe/H] $\sim-2.5$ the mean barium abundance is
$\langle$[Ba/Fe]$\rangle$ $\sim$ --1.8 $\pm$ 0.4, significantly
smaller than for {\boo}, for which $\langle$[Ba/Fe]$\rangle$ $\sim$
--0.5 $\pm$ 0.2 -- suggestive of a difference in the production
efficiency of the heavy-neutron-capture elements between the two
systems.  For UMa~II, on the other hand, one finds
$\langle$[Ba/Fe]$\rangle$ $\sim$ --1.0 $\pm$ 0.5, the same to within
the errors as that of {\boo}.

We conclude by noting that the {\boo} C-rich star (Boo-119) has a low
value of [Ba/Fe], signaling its membership of the CEMP-no class.  An
important area for future study is the determination of [Ba/Fe] in the
four other C-rich stars ([C/Fe] $>$ 0.7) in Figure~\ref{Fig:xfe} to
constrain more strongly the fraction of CEMP-no stars in the {\uf}
dwarf galaxies.
 
\section{DISCUSSION}\label{Sec:discussion}  

\subsection{The Properties of {\boo}}

{\boo} currently is the only ultra-faint satellite galaxy with a
well-defined metallicity distribution derived from medium-resolution
spectroscopic observations, plus a significant number of member stars
with high-resolution measured abundances of a range of elements,
including carbon, across the entire range of [Fe/H] from $\sim -3.5$
to $\sim -2$~dex. The wealth of data for {\boo} allows
one to trace chemical evolution at the lowest abundances, with
confidence that the stars formed in the same potential well, in a
self-enriching system. This is in contrast with the situation for the
sample of stars in the Galactic halo field, whose histories and
origins are unknown.

{\boo} has a broad metallicity distribution with a well-defined peak,
characterized by a mean iron abundance of $-2.6$~dex with dispersion
(standard deviation) of 0.4~dex \citep{norris10b, lai11}, encompassing
stars as iron-poor as $-3.7$~dex.  The existence of stars of very low
chemical abundances, and the wide range of abundances, make {\boo}
fully consistent with being a self-enriched galaxy that had an
essentially primordial initial metallicity.  The low value of the mean
stellar metallicity is consistent with supernovae-driven loss of
$\simgt 90\%$ of the initial baryons during star formation and
self-enrichment, assuming a normal IMF and standard nucleosynthetic
yields during the enrichment (cf. \citealp{hartwick76}), consistent
with the IMF inferred from the elemental abundances of the bulk of the
stars.  The color-magnitude diagram of {\boo} is that of a metal-poor,
exclusively old population \citep{belokurov06, okamoto12}, also
consistent with early star-formation truncated by gas loss.
 
The radial-velocity distribution of candidate member stars is offset
from that of most field stars, ensuring reliable membership
identification. The mass inferred from the kinematics is orders of
magnitude larger than the stellar mass (estimated to be $4 \times
10^4$M$_\odot$ for an assumed normal IMF \citep{martin08}), and
implies dark-matter domination. Scaling from the values given in
Table~1 of \citet{walker09} to take account of the revised velocity
dispersion value noted earlier, the dark-matter mass of {\boo} is
$\lesssim 10^7$M$_\odot$, and adopting the half light radius of $\sim
200$pc as a fiducial scalelength, the virial temperature
characterizing the potential well is $\sim 3 \times 10^3$~K: this sets
the initial temperature of gas, assuming it is shock-heated as it
comes into equilibrium within the dark-matter potential.  The high
mass-to-light ratio also implies that {\boo} lost $\simgt 90\%$ of its
baryons, if the initial value of baryonic to non-baryonic matter were
the cosmic value.

The mean mass density of {\boo}, inferred from its internal stellar
kinematics is $<\rho> \sim 0.025$~M$_\odot$~pc$^{-3}$, where this
value reflects a reduction of a factor of four from the mean density
derived in \citet{walker09}, to correct for the newer and lower value
of the central velocity dispersion determined by \citet{koposov11}.
Using this value, the simple assumption of dissipationless collapse of
a top-hat spherical density perturbation leads to an estimated
virialization redshift $\simgt 10$, consistent with {\boo} having
started star formation prior to the completion of reionization.

Although this is well-established as yet only for {\boo}, in general
as data improve the ultra-faint dwarf spheroidals are showing
abundance distribution functions and inferred stellar age
distributions consistent with being surviving examples of the first
systems to form stars (e.g.~\citealp{bovill11};
  \citealp{brown12}).  The ultra-faint galaxies therefore are of
critical importance in understanding early star formation and chemical
evolution. The massive stars in these systems could well be important
sources of the ionizing photons that contributed to reionization.

\subsection{Chemical Evolution of {\boo}: Implications from the $\alpha$-elements}

The $\alpha$-elements, together with a small amount of iron, are
created and ejected by core-collapse supernovae, on timescales of less
than $10^8$yr after formation of the supernova progenitors. Enhanced
(above solar) ratios of [$\alpha$/Fe] are expected in stars formed
from gas that is predominantly enriched by these end-points of massive
stars. Thus chemical abundances in the stars formed in the first
$\lesssim 0.5$~Gyr after star-formation began will reflect the
products of predominantly core-collapse supernovae.  As has already
been noted in the case of the field halo (see also \citealp{nissen94}
and \citealp{arnone05}), a lack of scatter in these element ratios at
given [Fe/H] requires that: (i) the stars formed from gas that was
enriched by ejecta sampling the mass range of the progenitors of
core-collapse supernovae; (ii) the supernova progenitor stars formed
with an IMF similar to that of the solar neighborhood today; and (iii)
the ejecta from all SNe were efficiently well-mixed.

Both the first and last points set an upper limit on how rapidly star
formation could have proceeded, since the star formation regions need
to populate the entire massive-star IMF, the stars need sufficient
time to all explode, and the gas needs time to mix the ejected
enriched material -- all before substantial numbers of low-mass stars
form.

Our formal full-sample result derived above is that the six
carbon-normal stars analyzed here show no evidence for any deviation
in the mean value of [$\alpha$/Fe] or any resolved scatter in element
ratios, within a limit of $\sim 0.1$~dex, with the exception of a
single star, Boo-41, with an anomalous abundance of a single element,
Ti. We noted the intriguing possibility that Boo-41 as an outlier star
is masking evidence of a resolved steady decline in mean [$\alpha$/Fe]
with [Fe/H]. This (possible) decline in the values of [$\alpha$/Fe],
as a function of [Fe/H], suggested by the data in
Figure~\ref{Fig:alpha8} could reflect the injection of iron from Type
Ia, indicating that the duration of star formation was a time
comparable to that for Type Ia SNe to explode and for their ejecta to
be incorporated in the next generation, $\lesssim$~1Gyr.  The apparent
decline could also simply reflect small number statistics, as noted
earlier.

The upper limit in intrinsic scatter of the $\alpha$-element to iron
ratios, as a function of iron, that was derived for our six
carbon-normal stars in Section~\ref{Sec:dispersions}, implies that
mixing of the ISM was efficient across the scales probed by the
sample.  The knowledge that all the stars are members of {\boo} -- and
presumably formed there -- provides a critical piece of information
that cannot be ascertained for the field halo, namely the physical
scale over which mixing must be efficient. The present sample probes
projected distances from the center of {\boo} of several hundred
parsecs, and this is the natural scale to adopt.

Type Ia supernovae produce significant amounts of iron, about ten
times as much per supernova as core-collapse events, but contribute
little to the abundances of the alpha-elements.  The \lq Delay Time
Distribution' describing the time between formation of the progenitors
and the subsequent Type Ia supernova explosions is model dependent;
however this lag cannot be shorter than the lifetime of an 8~M$_\odot$
star ($\sim 10^8$yr). Popular models have a peak rate at intermediate
delays ($\sim 10^9$~yr), with non-negligible rates at delays of many
Gyr, continuing to very late times (e.g. Figure~1 of
\citealp{matteucci09}).  The iron abundance at which the
nucleosynthetic products from Type Ia supernovae are manifest in the
next generation of stars depends on the detailed gas physics and the
star formation rate, with lower rates allowing the signatures of Type
Ia -- lower values of [$\alpha$/Fe] -- to occur at a lower iron
abundance.

Star formation in extremely metal-poor gas in low-mass halos (virial
temperature less than $10^4$~K so that cooling by atomic hydrogen is
not feasible) depends critically on the ionized fraction, as this
controls the creation of both molecular hydrogen and the HD molecule,
both of which can provide efficient cooling channels
(e.g. \citealp{johnson06}).  Suffice to say that the cooling is
complex, but a low gas temperature may be expected, $T \lesssim 100
$~K, with corresponding sound speed in atomic hydrogen of $\sim
2$~{\kms} (similar to the observed stellar velocity dispersion).  Star
formation in such systems has been shown to lead to a metal-poor
population of stars with a mass function encompassing low masses
\citep{clark11}.  The sound-crossing time may be taken as a
characteristic timescale for transport of metals and thus a limit on
the mixing timescale.  Mixing over the half-light radius of {\boo}
then took of order $\sim 10^8$~yr.  The lack of intrinsic scatter in
the $\alpha$-element ratios of stars within the half-light radius then
requires a minimum duration of star formation in {\boo} of $\gtrsim
10^8$~yr.  This is long enough that the progenitor-mass range of
core-collapse supernovae should be fully sampled and some Type Ia
supernovae may have occurred.

As noted above, the present stellar mass of {\boo} is $\sim 4 \times
10^4$~M$_\odot$; a normal stellar IMF implies an initial mass of stars
of around a factor of two higher (a locked-up fraction of around 50\%,
long after star formation ceased). The time-averaged star-formation
rate, using the duration of $10^8$~yr from above, is then less than
0.001~M$_\odot$/yr. This is small in absolute terms, but the
one-dimensional velocity dispersion of {\boo} is only $\sim 3~${\kms},
well below the critical minimum value for retention of gas in
idealized models of supernova feedback from star-formation on a
crossing time \citep{wyse85}.

A lack of scatter in the $\alpha$-elemental abundance ratios also
requires that the well-sampled IMF of core-collapse supernova
progenitors be invariant over the range of time and/or iron abundance.
As discussed by \citet{wyse92} and \citet{nissen94}, and revisited in
\citet{ruchti11}, the value of the ratios of the $\alpha$-elements to
iron during the regimes where core-collapse supernovae dominate the
chemical enrichment is a sensitive measure of the masses of the
progenitors of the supernovae. This is most simply expressed as a
constraint, from the scatter, on the variation in slope of the
massive-star IMF, assuming that the ratios reflect IMF-averaged
yields.  A scatter of $\pm 0.02$ constrains the variation in IMF slope
to be $\pm 0.2$ \citep[their Figure~30]{ruchti11}. The overall
agreement between the values of the elemental abundances in {\boo}
stars and in the field halo implies the same value of the massive-star
IMF that enriched the stars in each of the two samples, though our
formal limit on this IMF slope is only agreement within a slope range
of $\pm1$.

\subsection{Chemical Evolution of {\boo} and CEMP Stars: Implications from Carbon Abundances}

{\boo} and the Galactic halo display a large spread in carbon
abundance at lowest metallicity.  At [Fe/H] = --3.5, for example, both
exhibit a range of $\Delta$[C/Fe]$\sim2 - 3$ dex (see, e.g., the
top-left panel of Figure~\ref{Fig:xfe}).  Further, the fraction of
CEMP stars in {\boo} relative to that of giants in the halo in the
range --4 $<$ [Fe/H] $<$ --2 is $\sim$~0.2, similar to that found in
the Galactic halo.  That said, while the halo CEMP class comprises
several subclasses (CEMP-r, -r/s, -s, -no; see the classification of
\citealp{beers05}) little is known about the situation for the dwarf
spheroidal and {\uf} satellites.  As discussed in
Section~\ref{Sec:mgca}, in these systems data of quality sufficient to
determine their subclass exist for only two CEMP stars.  These are
Boo-119 and Segue 1-7, both of which belong to the CEMP-no subclass.
Both also have [Fe/H] $\sim-3.5$, consistent with the finding that in
the Galactic halo only CEMP-no stars exist below [Fe/H[ $\sim$ --3.2
    (see, e.g., \citealp{aoki10} and \citealp{norris12}).

It should be noted that the most iron-poor star currently known in
{\boo}, {\boos}, is not carbon-enhanced. Carbon-enhanced and
``carbon-normal'' stars co-exist at the same low iron abundance within
the same system. The recent discovery by \citet{caffau11, caffau12} of
an ultra metal-poor ([Fe/H]$_{\rm 1D,LTE}$ = --4.7, [Fe/H]$_{\rm
  3D,NLTE}$ = --4.9) low-mass star, in the Milky Way halo with no
carbon enhancement ([C/Fe] $<$ +0.9) is further evidence that carbon
enhancement is not required for very low-iron abundance gas to cool
and form low-mass stars.

CEMP stars that exhibit high values of neutron-capture $s-$process
elements (the CEMP-s stars) show a high incidence of binarity
(e.g. \citealp{lucatello05}) and are probably carbon-rich due to the
accretion of enriched material from an asymptotic giant branch
companion.  As shown by \citet{norris12}, however, the binary
statistics for CEMP-no stars are decidedly different from those of
CEMP-s stars, offering little support, currently, for a large
percentage of CEMP-no stars belonging to binary systems.  CEMP-no
stars are more likely to have formed from gas enriched by non-standard
(``mixing and fallback'' supernovae \citep{umeda03, iwamoto05}, or by
the winds from rapidly-rotating massive stars \citep{meynet06,
  meynet10}.  Unusually elevated magnesium is frequently also found in
such stars \citep{aoki02, norris12}, with presumably both the
magnesium and carbon over-abundances reflecting the yields of the very
first generation of supernovae or massive stars. The presence of
CEMP-no stars in the two {\uf} dwarf galaxies is strong evidence for
their self-enrichment from primordial material. It provides an
opportunity to consider the evolutionary history of the extremely
carbon-enriched, Fe-poor ISM gas.

The important additional information we consider here is that CEMP-no
stars are not found at [Fe/H]~$\simgt -2.5$, either in the field halo
\citep{aoki10} or in dwarf spheroidal galaxies \citep{norris10b,
  lai11}\footnote {Note, however, that only very meager information
  exists concerning the sub-classification of CEMP stars in dSph
  galaxies.  To our knowledge all that is currently known is that
  Boo-119 (\citealp{lai11}; this work) and Segue 1-7 \citep{norris10c}
  are both CEMP-no stars.}.  Given the amplitude of the [C/Fe] and
      [Mg/Fe] values in CEMP-no stars, one must also explain why stars
      are not found with intermediate C and Mg excesses at higher
      [Fe/H]. Apparently the highly C- and Mg-enriched ISM does not
      survive to mix with ``normal'' enriched ISM and form more stars
      with moderate CEMP-no enrichment. Rather, the cooling efficiency
      of the highly carbon-enriched material must be sufficiently
      great that all of it cools and forms (the surviving) low-mass
      stars before there is time to mix this material with ``normal''
      SNe ejecta.

\begin{figure}[htbp]
\begin{center}
\includegraphics[width=6.5cm,angle=270]{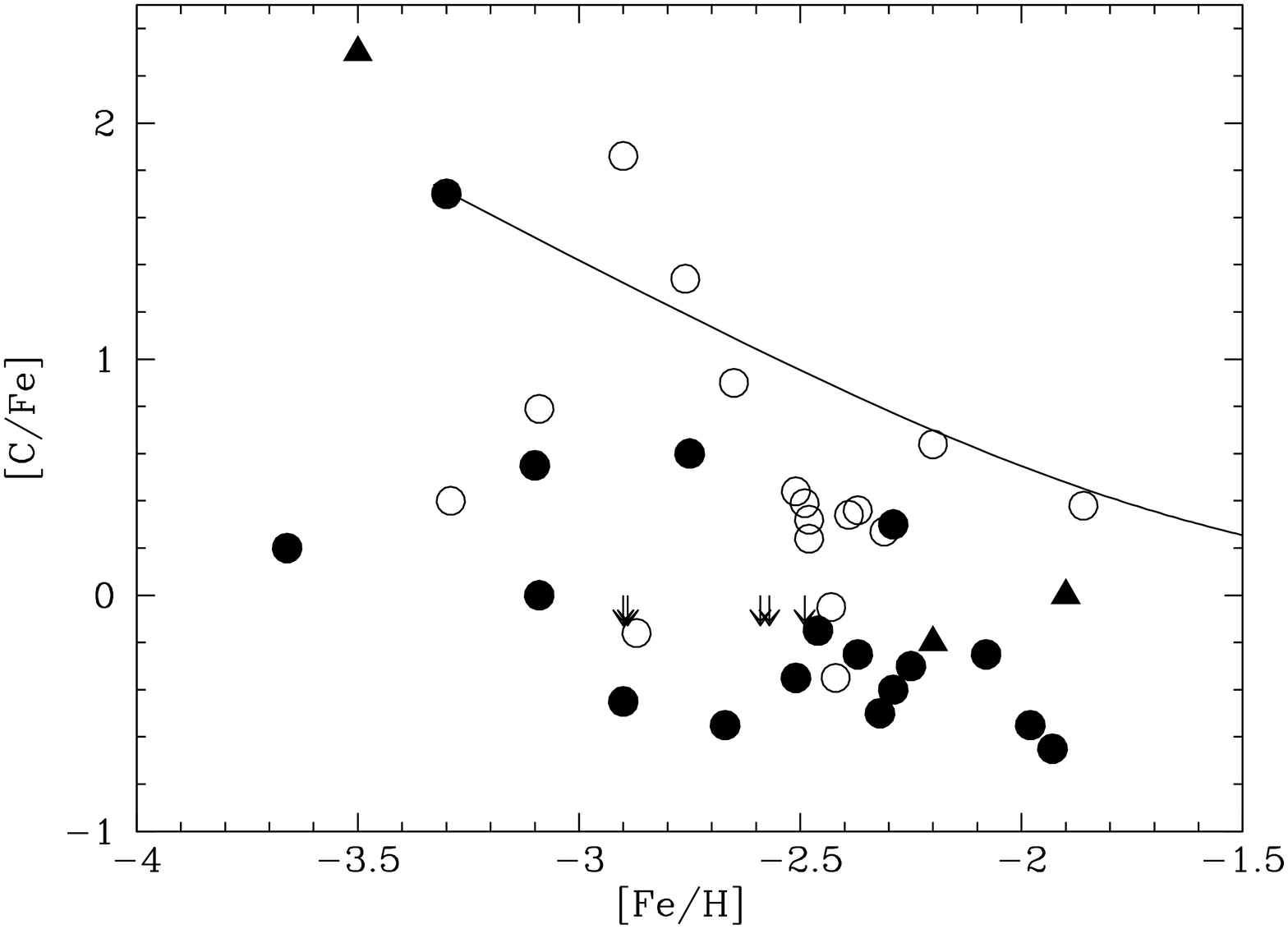}

  \caption{\label{Fig:rw_cfe} [C/Fe] vs. [Fe/H] for stars in the
    {\boo} dwarf galaxy (filled circles represent data of
    \citet{norris10b}, with the exception of the CEMP-no star Boo-119,
    which has been plotted adopting the carbon abundance from
    \citet{lai11} and the iron abundance from the present work; open
    circles and upper limits represent data from \citet{lai11}) and
    for the Segue~1 dwarf galaxy (filled triangles, data from
    \citet{norris10b,norris10c}). The smooth curve starts at Boo-119,
    and tracks the expected evolution due to continuing star formation
    in gas to which carbon and iron are added in the solar ratio,
    consistent with normal core-collapse supernovae. As discussed in
    the text, the data suggest two distinct channels of enrichment,
    one carbon-rich, and one carbon-normal.}

\end{center}
\end{figure}

This is illustrated in Figure~\ref{Fig:rw_cfe}, which shows [C/Fe]
vs. [Fe/H] for stars in the {\boo} dwarf galaxy, based on two samples:
filled circles represent data of \citet{norris10a}, with the exception
of the CEMP-no star Boo-119, which has been plotted adopting the
carbon abundance from \citet{lai11} and the iron abundance from the
present work, while open circles and upper limits represent data from
\citet{lai11}.  The smooth curve starts at Boo-119, and tracks the
expected evolution due to addition of enriched gas in which carbon and
iron are in the solar ratio, consistent with normal core-collapse
supernovae and the value of the [C/Fe] ratio for the bulk of field
halo stars. This curve skims along the top of the distribution for
{\boo} and clearly fails to explain the bulk of the ``carbon-normal''
population.  This suggestion is supported also by the data for the
Segue 1 {\uf} system, for which the available sample having carbon
abundances is limited to three stars \citep{norris10b,norris10c},
shown as filled triangles in Figure~\ref{Fig:rw_cfe}. The analogous
carbon-dilution curve for Segue~1, starting at the CEMP-no star Segue
1-7 at ([C/Fe], [Fe/H]) = (+2.3, --3.5) \citep{norris10c}, would also
fail to explain the carbon-normal stars in this system. We conclude
that there appears to be two distinct channels of enrichment, one
creating CEMP-no stars, and the other creating carbon-normal stars.
Both channels operate at low iron abundance, [Fe/H] $\lesssim -2.5$,
but only the carbon-normal channel remains at higher
[Fe/H]. \citet{norris12} reached the same conclusion from their
analysis of the abundance patterns of Galactic halo C-rich stars with
[Fe/H] $<$ --3.1.

\begin{figure}[htbp]
\begin{center}
\includegraphics[width=8.0cm,angle=0]{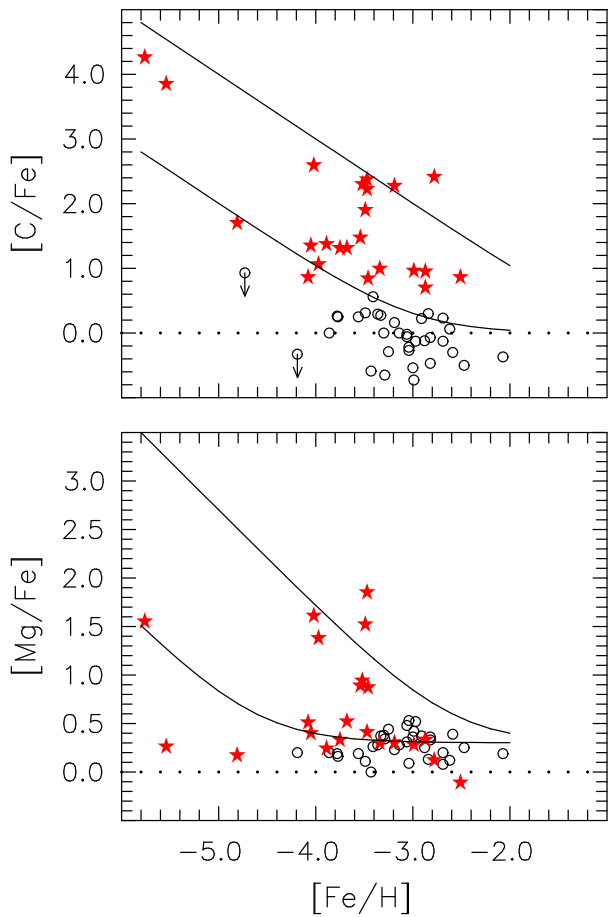}

\caption{\label{Fig:cmgfe} [C/Fe] and [Mg/Fe] vs. [Fe/H] for Galactic
  halo CEMP-no stars (star symbols, from Norris et al.\ 2012) and
  C-normal stars (circles, from \citealp{cayrel04} and
  \citealp{caffau11}). The lines represent dilution trajectories, as
  C-rich and Mg-rich material produced at the earliest times is mixed
  with C-normal and Mg-normal halo material.  As [Fe/H]
  increases to --2.5, early C-rich and Mg-rich signatures
  are no longer evident.  See text for discussion. }

\end{center}
\end{figure}

The inability to form the bulk of the carbon-normal stars by an
earlier population of CEMP-no stars is most clear in the case of
individual self-enriching systems, as opposed to a sample of stars in
the field halo with unknown, and likely varied, formation sites.  The
available sample of field halo stars is shown in
Figure~\ref{Fig:cmgfe}, where the analysis has been extended to
include [Mg/Fe] (see \citealp{norris12} for a full description of
these data). Dilution tracks can be found that start at observed
values and pass through the bulk of the stars. However, there is no
guarantee that stars so-connected were ever part of the same
self-enriching system.

A comparison between the distributions and carbon-dilution tracks in
Figures~\ref{Fig:rw_cfe} and \ref{Fig:cmgfe} shows the power of
knowing that the stars in the {\boo} sample are part of a
self-enriching system and therefore can be connected by a chemical
evolutionary path: choosing a different starting point for the
carbon-dilution track in Figure~\ref{Fig:rw_cfe} could provide values
of [C/Fe] and [Fe/H] that better match those of the bulk of the stars.
Indeed, material with an initial [C/Fe] of the same value as for
Boo-119, but a factor of a hundred lower iron abundance, would, after
dilution by material with solar carbon-to-iron, pass through the locus
of the ``carbon-normal'' stars in {\boo} at [Fe/H] $ \simgt -3$. The
challenge would be to find the evidence that a sufficient population
of such precursers existed. We can robustly state that with the
current knowledge of stars in {\boo} there is no direct chemical
evolution track (assuming standard yields of carbon and iron) between
the CEMP-no star and the ``carbon-normal'' stars.

\begin{figure*}[htbp]
\begin{center}
\includegraphics[width=12.0cm,angle=0]{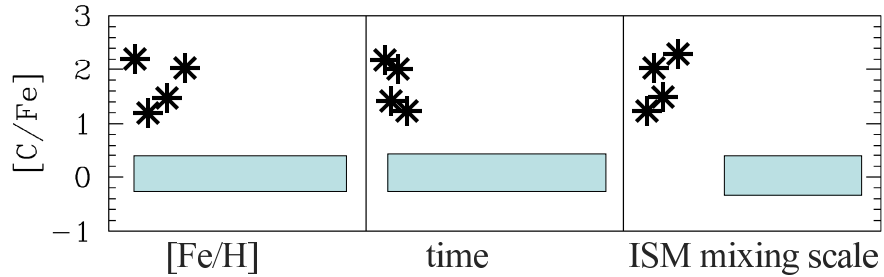}

  \caption{\label{Fig:carb-evol}Schematic illustration of (i) (left
    panel) the different iron abundance regimes of the two
    carbon-enrichment channels; (ii) (middle panel) the different
    timescales since the onset of star-formation applying to the two
    separate enrichment channels and (iii) (right panel) the two
    different spatial scales over which nucleosynthetic material was
    mixed. The star symbols indicate the CEMP-no channel, while the
    blue band indicates the carbon-normal channel.}

\end{center}
\end{figure*}

A schematic diagram of the processes underlying the [C/Fe] patterns
associated with these two paths is given in
Figure~\ref{Fig:carb-evol}.  This shows, for illustrative purposes,
four stars (or star-forming events) in the ``CEMP-no'' path, denoted
by star symbols, while the ``normal'' path lies along the blue shaded
area.  The three panels indicate different aspects of enrichment in
which we propose that the two paths diverge: the first is [C/Fe]
versus [Fe/H], just discussed above; the second is [C/Fe] versus time
since the onset of star formation, indicating that star formation of
``CEMP-no'' stars is initiated prior to the ``normal'' branch, but may
overlap somewhat in time before disappearing; and the third shows
[C/Fe] versus mixing length in the ISM.  The CEMP-no stars form
rapidly out of gas enriched by only one generation of supernovae and
likely prior to the onset of good mixing.  This results in a small
mixing length, spatial inhomogeneity and a large scatter in elemental
abundance ratios.

This scenario requires that the gas within which the CEMP-no stars
form can cool and be locked up in low-mass stars very rapidly, and
with high efficiency, so that material with this abundance pattern is
removed from the system at early times. Several models of the
formation of very metal-poor low-mass stars have appealed to enhanced
cooling due to carbon (e.g. review of \citealp{bromm04}).  It may well
be that the CEMP-no material resulted from a very small number of
(Population III?) supernovae.

\section{SUMMARY}

We have analyzed UVES spectra of seven red giant members of {\boo},
and re-analyzed our previous UVES study of an eighth member. The stars
cover the range in [Fe/H] from --3.7 to --1.9, and include a CEMP-no
star with [Fe/H]= --3.33.

We implemented a double-blind analysis strategy to ensure the most
reliable feasible determination of the measuring errors with which
elemental abundances could be determined.  We did this since the
analysis of elemental abundances in low-mass stars in ultra-faint
dwarf galaxies provides unique insight into early star formation and
chemical enrichment. These systems are inferred to be very dark-matter
dominated, are of extremely low surface brightness, with low stellar
mass, and have a uniformly old stellar population. 

Our elemental abundances are formally consistent with a halo-like
distribution, with enhanced mean [$\alpha$/Fe] and at most small
scatter about the mean. This is in accord with the high-mass
low-metallicity stellar IMF in this very low density system being
indistinguishable from the present-day solar neighborhood value.  We
do find one star with apparently very high [Ti/Fe] abundance, while we
also find no support for a previously published high [Mg/Fe] value for
another of our stars.  We see marginal hints of a decline in
[$\alpha$/Fe] with [Fe/H] if we exclude the high-Ti abundance star.
Further observations are needed to examine this tentative result.

Our metallicity and elemental abundance data show that {\boo} has
evolved as a self-enriching star-forming system, from essentially
primordial initial abundances. This allows us uniquely to investigate
the place of CEMP-no stars in a chemically evolving system, as well as
to limit the timescale of star formation in this {\uf} dSph. Both the
low elemental abundance scatter and the hint of a decline in
[$\alpha$/Fe] require low star formation rates, allowing time for SNe
ejecta to be created and mixed over the large spatial scales relevant
to {\boo}. This is further evidence that {\boo} survived as a
self-enriching star-forming system from very early times.

In that context we consider the implications of the existence of
CEMP-no stars at very low values of [Fe/H], and the absence of any
stars with lesser CEMP-no enhancements at higher [Fe/H] -- a situation
which is consistent with knowledge of CEMP-no stars in the Galactic
field.  We show that this observation requires there are two
enrichment paths at very low metallicities: CEMP-no and
``carbon-normal''.

\acknowledgements

Studies at RSAA, ANU, of the Galaxy's most metal-poor stars and
ultra-faint satellite systems are supported by Australian Research
Council grants DP0663562 and DP0984924, which J.E.N. and D.Y. are
pleased to acknowledge.
R.F.G.W. acknowledges partial support from the US National Science
Foundation through grants AST-0908326 and CDI-1124403, and thanks the
Aspen Center for Physics (supported by NSF grant PHY-1066293) for
hospitality while this work was completed.

\noindent{\it Facilities:} {VLT:Kueyen(UVES)}

\end{document}

%% file: tab1.tex
\begin{deluxetable*}{rcccccccc}
\tablecolumns{9} 
\tablewidth{0pt}
\tablecaption{\label{Tab:program}THE SEVEN BO\"{O}TES I RED-GIANT PROGRAM STARS}
\tablehead{
\colhead {Star} & {Other} & {RA}        & {Dec}    & {$g_{0}$}  & {$(g-r)_{0}$}   & {[Fe/H]\tablenotemark{b}} & {[Fe/H]\tablenotemark{b}}& {[Fe/H]\tablenotemark{b}}\\
         {}     &    {ID\tablenotemark{a}} & {(2000)}    & {(2000)} & {}  & {}    & {}  & {} & {}             \\ 
         {(1)}  & {(2)}       & {(3)}    & {(4)}      & {(5)}           & {(6)} & {(7)}  & {(8)}  & {(9)}     
}
\startdata
33  &  ... ... & 14 00 11.73 & +14 25 01.4 & 18.155 & 0.736 & --2.96 & ...  & ... \\   
41  &   66  29 & 14 00 25.83 & +14 26 07.6 & 18.304 & 0.697 & --2.03 & --1.6  & --1.65  \\   
94  &  ... ... & 14 00 31.51 & +14 34 03.6 & 17.449 & 0.872 & --2.79 & ...  & ...  \\   
117 &    4   1 & 14 00 10.49 & +14 31 45.5 & 18.134 & 0.746 & --1.72 & --2.2  & --2.34  \\   
119 &   63  21 & 14 00 09.85 & +14 28 22.9 & 18.359 & 0.728 &    ... & --2.7  & --3.79 \\
127 &  ... ... & 14 00 14.57 & +14 35 52.7 & 18.087 & 0.773 & --1.49 & ...  & ...  \\   
130 &  ... ... & 13 59 48.98 & +14 30 06.2 & 18.136 & 0.707 & --2.55 & ...  & ...     
\enddata
\tablenotetext{a} {Identifications of \citet[row number of their Table 1]{martin07} and \citet{lai11}, respectively}
\tablenotetext{b} {Determined by \citet{norris08}, \citet{martin07} and \citet{lai11} using the Ca II K line, the Ca II infrared triplet, and intermediate-resolution blue data, respectively
\vspace{3mm} }
\end{deluxetable*}

%% file: tab2.tex
\begin{deluxetable*}{rrccrccr}
\tablecolumns{8} 
\tablewidth{0pt}
\tablecaption{\label{Tab:radvels}RADIAL VELOCITIES FOR PROGRAM STARS}
\tablehead{
\colhead {Star} & {V$_{\rm r}$(NY)} & {s.e.V$_{\rm r}$(NY)} & {No.} & {V$_{\rm r}$(GM)} & {s.e.V$_{\rm r}$(GM)} & {No.}  & {V$_{\rm r}$}  \\
         {}     & {({\kms})}  & {({\kms})}             & {}    & {({\kms})} & {({\kms})}         & {}     & {({\kms})}  \\ 
         {(1)}  & {(2)}     & {(3)}                    & {(4)} & {(5)} & {(6)}                   & {(7)}  & {(8)}     
}
\startdata
    33   &  102.11  &  0.31  &  21  &  102.18  &  0.08  &  22  &    102.15 \\
    41   &  106.91  &  0.29  &  21  &  106.97  &  0.10  &  22  &    106.94 \\
    94   &   95.34  &  0.28  &  21  &   95.51  &  0.14  &  22  &     95.43 \\
    117  &   99.45  &  0.30  &  21  &   99.64  &  0.08  &  22  &     99.54 \\
    119  &   91.75  &  0.30  &  20  &   93.25  &  0.33  &  22  &     92.50 \\
    127  &  100.28  &  0.31  &  21  &  100.24  &  0.09  &  22  &    100.26 \\
    130  &  104.85  &  0.27  &  21  &  104.60  &  0.10  &  22  &    104.72 
\enddata

\end{deluxetable*}

%% file: tab3.tex
\begin{deluxetable*}{rcccccccccccc}
\tablecolumns{13} 
\tablewidth{0pt}
\tabletypesize{\scriptsize}
\tablecaption{\label{Tab:param}ATMOSPHERIC PARAMETERS FOR PROGRAM STARS}
\tablehead{
\colhead  {Star} & {\teff(K)} & {\logg} & {[Fe/H]\tablenotemark{a}} & {\vxi} &{\teff(K)} & {\logg} & {[Fe/H]\tablenotemark{a}} & {\vxi} &{\teff(K)}  & {\logg} & {[Fe/H]\tablenotemark{a}} & {\vxi} \\
          {}  & {(NY)}   & {(NY)}   & {(NY)}    & {(NY)}  & {(GM)}  & {(GM)}   & {(GM)}    & {(GM)}  &{(F09)}   & {(F09)}  & {(F09)}   & {(F09)} \\    
          {(1)}  & {(2)}   & {(3)}   & {(4)}    & {(5)}  & {(6)}  & {(7)}   & {(8)}    & {(9)}  &{(10)}   & {(11)}  & {(12)}   & {(13)}     
}
\startdata
33   &4730 &1.4 &--2.36 & 2.8  &4750 & 1.4 &--2.28 & 2.0  &4600 & 1.0 &--2.52 & 2.1 \\ 
41   &4750 &1.6 &--1.96 & 2.8  &4800 & 1.4 &--1.80 & 2.1  & ... & ... &  ...  & ... \\ 
94   &4570 &0.8 &--2.97 & 3.3  &4550 & 0.9 &--2.91 & 2.0  &$<$4600 & 0.5 &--2.95 & 2.1 \\ 
117  &4700 &1.4 &--2.31 & 2.7  &4750 & 1.4 &--2.05 & 1.8  &4600 & 1.0 &--2.29 & 2.1 \\ 
119  &4790 &1.4 &--3.21 & 2.4  &4750 & 1.4 &--3.44 & 2.9  & ... & 1.0 &  ...  & ... \\ 
127  &4670 &1.4 &--2.11 & 2.7  &4700 & 1.4 &--1.92 & 2.0  &4600 & 1.0 &--2.03 & 2.1 \\ 
130  &4750 &1.4 &--2.35 & 2.6  &4800 & 1.4 &--2.28 & 2.1  & ... & 1.0 &  ...  & ...  
\enddata
\tablenotetext{a} {Assuming log\,$\epsilon_{\odot}$(Fe) = 7.50, following \citet{asplund09}}
\end{deluxetable*}

%% file: tab4e.tex
\begin{deluxetable*}{lccrrrrrrrrrrr}
\tabletypesize{\tiny}                                                                                                   
\tablecolumns{14}                                                                                                       
\tablewidth{0pt}                                                                                                        
\tablecaption{\label{Tab:nyeweps}NY ATOMIC DATA, EQUIVALENT WIDTHS (m{\AA}), AND ABSOLUTE ABUNDANCES}
\tablehead{
\colhead {Spec.} & {$\lambda$} & {$\chi$} & {\loggf} & {EW} & {log\,$\epsilon$} & {EW} & {log\,$\epsilon$} & {EW} & {log\,$\epsilon$} & {EW} & {log\,$\epsilon$} & {EW} & {log\,$\epsilon$}  \\
          {}  &   {(\AA)}      &  {(eV)} &  {}     & {Boo33} & {Boo33}      & {Boo41} &   {Boo41}    & {Boo94} & {Boo94}   & {Boo117} & {Boo117}     & {Boo119} & {Boo119}  \\
({1}) & ({2}) & ({3}) & ({4}) & ({5}) & ({6}) & ({7}) & ({8}) & ({9}) & ({10}) & ({11}) & ({12}) & ({13}) & ({14}) 
}                                                                                                                       
\startdata                                                                                                              
Na\,I    &  5895.92 &   0.00 &  $-$0.19 & 191.0 &  3.90 &   ... &   ... & 164.5 &  3.22 &   ... &   ... & 155.0 &  3.73  \\
Mg\,I    &  5528.40 &   4.34 &  $-$0.34 & 108.0 &  5.42 & 168.5 &  6.16 &  81.3 &  4.98 & 104.5 &  5.36 &  90.7 &  5.33  \\
Ca\,I    &  5349.47 &   2.71 &  $-$0.31 &  17.5 &  4.00 &  59.8 &  4.70 &   ... &   ... &  31.0 &  4.28 &   ... &   ...  \\
Ca\,I    &  5581.98 &   2.52 &  $-$0.71 &   ... &   ... &  46.3 &  4.69 &   ... &   ... &  37.2 &  4.55 &   ... &   ...  \\
Ca\,I    &  5588.75 &   2.52 &   0.21   &  82.4 &  4.27 & 118.5 &  4.72 &   ... &   ... &  83.6 &  4.27 &  18.0 &  3.32  
\enddata                                                                                                                

\tablerefs{Notes. 
~
(This table is available in its entirety in a machine-readable
form in the online journal. A portion is shown here for guidance regarding its
form and content.)} 

\end{deluxetable*}

%% file: tab5e.tex
\begin{deluxetable*}{lccrrrrrrrrrrr}
\tabletypesize{\tiny}                                                                                                   
\tablecolumns{14}                                                                                                       
\tablewidth{0pt}                                                                                                        
\tablecaption{\label{Tab:gmeweps}GM ATOMIC DATA, EQUIVALENT WIDTHS (m{\AA}), AND ABSOLUTE ABUNDANCES}
\tablehead{
\colhead {Spec.} & {$\lambda$} & {$\chi$} & {\loggf} & {EW} & {log\,$\epsilon$} & {EW} & {log\,$\epsilon$} & {EW} & {log\,$\epsilon$} & {EW} & {log\,$\epsilon$} & {EW} & {log\,$\epsilon$}  \\
          {}     &   {(\AA)}   &  {(eV)}  &     {} & {Boo33} & {Boo33}      & {Boo41} &   {Boo41}    & {Boo94} & {Boo94}   & {Boo117} & {Boo117}     & {Boo119} & {Boo119}      \\
({1}) & ({2}) & ({3}) & ({4}) & ({5}) & ({6}) & ({7}) & ({8}) & ({9}) & ({10}) & ({11}) & ({12}) & ({13}) & ({14}) 
}                                                                                                                       
\startdata                                                                                                              
%            atomic                      |      33       |      41       |      94       |        117    |     119       |      127      |        130    |

   Na\,I &  5895.924 &   0.000 &  $-$0.184 & 177.6 &  4.04 &   ... &   ... & 154.6 &  3.68 & 177.2 &  4.14 & 167.8 &  3.55  \\
   Mg\,I &  5528.405 &   4.346 &  $-$0.620 &  93.1 &  5.66 &   ... &   ... &  75.4 &  5.36 & 101.9 &  5.85 &  74.6 &  5.30  \\
   Mg\,I &  5711.088 &   4.346 &  $-$1.833 &   ... &   ... &  49.5 &  6.27 &  10.2 &  5.28 &   ... &   ... &   ... &   ...  \\
   Si\,I &  5948.541 &   5.082 &  $-$0.780 &   ... &   ... &  26.0 &  5.58 &   ... &   ... &  25.8 &  5.56 &   ... &   ...  \\
   Ca\,I &  5261.704 &   2.521 &  $-$0.579 &  27.8 &  4.34 &   ... &   ... &   ... &   ... &   ... &   ... &   ... &   ...  
\enddata                                                                                                  

\tablerefs{Notes. 
~
(This table is available in its entirety in a machine-readable
form in the online journal. A portion is shown here for guidance regarding its
form and content.)} 
             
\end{deluxetable*}

%% file: tab6.tex
\begin{deluxetable*}{lrrrrrrrrrrr} 
\tablecolumns{12}
\tabletypesize{\tiny}                                                                                                   
\tablewidth{0pc} 
\tablecaption{\label{Tab:xfe}1D LTE ABUNDANCES OF THE FIRST FOUR BO\"{O}TES I GIANTS} 
\tablenum{6} 
\tablehead{ 
\colhead {Species} & {log\,$\epsilon_{\odot}$}& {log\,$\epsilon$}  & {s.e.$_{\log\epsilon}$}    & {N}   & {[X/Fe]} & {$\sigma$[X/Fe]} & {log\,$\epsilon$} & {s.e$_{\log\epsilon}$}     & {N}   & {[X/Fe]} & {$\sigma$[X/Fe]}\\
                 &                         & (NY)               & (NY)      & (NY)  & (NY)     & (NY)    & (GM)              & (GM)       & (GM)  & (GM)     & (GM)\\
            ({1})  & ({2})              & ({3})     & ({4}) & ({5})    & ({6})   & ({7})             & ({8})      & ({9}) & ({10})   & ({11}) & ({12})
}
\startdata 
 & & & & & & & & & & & \\                 
Boo-33  & & & & & & & & & & \\          
Na\,I  & 6.24 &    3.90 &      ... &      1 &    0.02 &    0.23     &    4.04 &      ... &      1 &    0.08 &    0.23 \\ 
Mg\,I  & 7.60 &    5.42 &      ... &      1 &    0.18 &    0.22     &    5.66 &      ... &      1 &    0.34 &    0.21 \\ 
Si\,I  & 7.51 &     ... &     ...  &     .. &     ... &     ...     &     ... &     ...  &     .. &     ... &     ... \\ 
Ca\,I  & 6.34 &    4.14 &    0.06  &      8 &    0.16 &    0.07     &    4.19 &    0.04  &     13 &    0.13 &    0.06 \\ 
Sc\,II & 3.15 &    0.64 &    0.08  &      3 &   $-$0.15 &    0.21     &    0.73 &    0.10  &      3 &   $-$0.14 &    0.21 \\ 
Ti\,I  & 4.95 &    2.51 &    0.12  &      5 &   $-$0.07 &    0.14     &    2.55 &    0.07  &      5 &   $-$0.11 &    0.09 \\ 
Ti\,II & 4.95 &    2.65 &    0.17  &      3 &    0.06 &    0.25     &    2.73 &    0.08  &      5 &    0.06 &    0.20 \\ 
Cr\,I  & 5.64 &    2.98 &    0.10  &      5 &   $-$0.29 &    0.11     &    3.13 &    0.04  &      5 &   $-$0.23 &    0.06 \\ 
Fe\,I\tablenotemark{a}  & 7.50  &    5.14 &    0.03  &     61 &   $-$2.36 &    0.16     &    5.22 &    0.03  &     51 &   $-$2.28 &    0.16 \\ 
Fe\,II & 7.50 &    5.11 &    0.13  &      4 &   $-$0.02 &    0.25     &    5.49 &    0.07  &      4 &    0.26 &    0.22 \\ 
Ni\,I  & 6.22 &    3.72 &    0.10  &      3 &   $-$0.14 &    0.15     &    4.01 &    0.08  &      2 &    0.07 &    0.10 \\ 
Zn\,I  & 4.56 &     ... &     ...  &     .. &     ... &     ...     &     ... &     ...  &     .. &     ... &     ... \\ 
Y\,II  & 2.21 &     ... &     ...  &     .. &     ... &     ...     &   $-$0.39 &      ... &      1 &   $-$0.32 &    0.22 \\ 
Ba\,II & 2.18 &   $-$0.67 &    0.09  &      4 &   $-$0.49 &    0.19     &   $-$0.40 &    0.04  &      2 &   $-$0.30 &    0.19 \\ 
 & & & & & & & & & & & \\                 
Boo-41  & & & & & & & & & & & \\          
Na\,I  & 6.24  &     ... &     ...  &     .. &     ... &     ...     &     ... &     ...  &     .. &     ... &     ... \\ 
Mg\,I  & 7.60  &    6.16 &      ... &      1 &    0.52 &    0.21     &    6.27 &      ... &      1 &    0.47 &    0.24 \\ 
Si\,I  & 7.51 &     ... &     ...  &     .. &     ... &     ...     &    5.58 &      ... &      1 &   $-$0.13 &    0.27 \\ 
Ca\,I  & 6.34 &    4.73 &    0.05  &     11 &    0.36 &    0.07     &    4.76 &    0.05  &      4 &    0.22 &    0.06 \\ 
Sc\,II & 3.15 &     ... &     ...  &     .. &     ... &     ...     &    1.10 &    0.10  &      5 &   $-$0.25 &    0.21 \\ 
Ti\,I  & 4.95 &    3.74 &    0.07  &      2 &    0.76 &    0.14     &    3.79 &    0.05  &      6 &    0.65 &    0.08 \\ 
Ti\,II & 4.95 &    4.03 &    0.32  &      2 &    1.05 &    0.37     &    3.85 &    0.07  &      3 &    0.70 &    0.19 \\ 
Cr\,I  & 5.64 &    3.74 &    0.08  &      5 &    0.06 &    0.09     &    4.23 &    0.03  &      3 &    0.39 &    0.09 \\ 
Fe\,I\tablenotemark{a}  & 7.50  &    5.54 &    0.03  &     47 &   $-$1.96 &    0.16     &    5.70 &    0.04  &     43 &   $-$1.80 &    0.16 \\ 
Fe\,II & 7.50  &    5.80 &    0.24  &      3 &    0.26 &    0.32     &    5.80 &    0.09  &      2 &    0.09 &    0.27 \\ 
Ni\,I  & 6.22  &    3.94 &      ... &      1 &   $-$0.32 &    0.21     &    4.57 &    0.08  &      6 &    0.15 &    0.09 \\ 
Zn\,I  & 4.56 &    2.75 &      ... &      1 &    0.15 &    0.25     &    3.13 &      ... &      1 &    0.37 &    0.24 \\ 
Y\,II  & 2.21 &     ... &     ...  &     .. &     ... &     ...     &     ... &     ...  &     .. &     ... &     ... \\ 
Ba\,II & 2.18 &   $-$0.19 &    0.10  &      4 &   $-$0.41 &    0.19     &    0.01 &    0.01  &      2 &   $-$0.37 &    0.21 \\ 
 & & & & & & & & & & & \\                 
Boo-94  & & & & & & & & & & & \\          
Na\,I  & 6.24  &    3.22 &      ... &      1 &   $-$0.05 &    0.21     &    3.68 &      ... &      1 &    0.35 &    0.18 \\ 
Mg\,I  & 7.60  &    4.98 &      ... &      1 &    0.35 &    0.19     &    5.32 &    0.05  &      2 &    0.63 &    0.12 \\ 
Si\,I  & 7.51 &     ... &     ...  &     .. &     ... &     ...     &     ... &     ...  &     .. &     ... &     ... \\ 
Ca\,I  & 6.34 &    3.65 &    0.04  &      6 &    0.27 &    0.06     &    3.75 &    0.04  &     10 &    0.32 &    0.06 \\ 
Sc\,II & 3.15 &    0.25 &    0.03  &      2 &    0.07 &    0.20     &    0.39 &    0.03  &      5 &    0.15 &    0.19 \\ 
Ti\,I  & 4.95 &    2.21 &    0.05  &      6 &    0.24 &    0.08     &    2.36 &    0.03  &     10 &    0.32 &    0.07 \\ 
Ti\,II & 4.95 &    2.23 &    0.07  &      5 &    0.24 &    0.19     &    2.29 &    0.02  &      6 &    0.25 &    0.18 \\ 
Cr\,I  & 5.64 &    2.33 &    0.05  &      4 &   $-$0.34 &    0.06     &    2.40 &    0.04  &      4 &   $-$0.33 &    0.06 \\ 
Fe\,I\tablenotemark{a}  & 7.50  &    4.53 &    0.02  &     64 &   $-$2.97 &    0.16     &    4.59 &    0.02  &     46 &   $-$2.91 &    0.16 \\ 
Fe\,II & 7.50  &    4.42 &    0.26  &      2 &   $-$0.11 &    0.34     &     ... &     ...  &     .. &     ... &     ... \\ 
Ni\,I  & 6.22  &    3.25 &    0.17  &      3 &   $-$0.01 &    0.17     &    3.40 &    0.05  &      4 &    0.09 &    0.07 \\ 
Zn\,I  & 4.56 &     ... &     ...  &     .. &     ... &     ...     &    2.06 &      ... &      1 &    0.41 &    0.18 \\ 
Y\,II  & 2.21 &     ... &     ...  &     .. &     ... &     ...     &   $-$1.09 &      ... &      1 &   $-$0.39 &    0.20 \\ 
Ba\,II & 2.18 &   $-$1.77 &    0.14  &      3 &   $-$0.98 &    0.21     &   $-$1.62 &      ... &      1 &   $-$0.89 &    0.19 \\ 
 & & & & & & & & & & & \\                 
Boo-117 & & & & & & & & & & & \\          
Na\,I  & 6.24  &     ... &     ...  &     .. &     ... &     ...     &    4.14 &      ... &      1 &   $-$0.05 &    0.24 \\ 
Mg\,I  & 7.60  &    5.36 &      ... &      1 &    0.07 &    0.20     &    5.85 &      ... &      1 &    0.30 &    0.23 \\ 
Si\,I  & 7.51 &     ... &     ...  &     .. &     ... &     ...     &    5.56 &      ... &      1 &    0.10 &    0.26 \\ 
Ca\,I  & 6.34 &    4.28 &    0.05  &     11 &    0.24 &    0.06     &    4.44 &    0.05  &     11 &    0.15 &    0.07 \\ 
Sc\,II & 3.15 &    0.86 &    0.04  &      3 &    0.02 &    0.21     &    1.08 &    0.02  &      5 &   $-$0.02 &    0.18 \\ 
Ti\,I  & 4.95 &    2.75 &    0.05  &      6 &    0.12 &    0.07     &    3.05 &    0.07  &      9 &    0.16 &    0.09 \\ 
Ti\,II & 4.95 &    2.76 &    0.07  &      4 &    0.12 &    0.19     &    3.10 &    0.02  &      6 &    0.20 &    0.18 \\ 
Cr\,I  & 5.64 &    3.17 &    0.05  &      6 &   $-$0.16 &    0.07     &    3.44 &    0.07  &      5 &   $-$0.15 &    0.08 \\ 
Fe\,I\tablenotemark{a}  & 7.50  &    5.19 &    0.02  &     61 &   $-$2.31 &    0.16     &    5.45 &    0.03  &     57 &   $-$2.05 &    0.16 \\ 
Fe\,II & 7.50  &    5.22 &    0.26  &      3 &    0.03 &    0.33     &    5.51 &    0.04  &      4 &    0.06 &    0.22 \\ 
Ni\,I  & 6.22  &    3.89 &    0.16  &      3 &   $-$0.03 &    0.17     &    4.12 &    0.04  &      9 &   $-$0.05 &    0.06 \\ 
Zn\,I  & 4.56 &    2.53 &      ... &      1 &    0.28 &    0.24     &    2.64 &      ... &      1 &    0.13 &    0.19 \\ 
Y\,II  & 2.21 &     ... &     ...  &     .. &     ... &     ...     &   $-$0.35 &    0.12  &      2 &   $-$0.52 &    0.21 \\ 
Ba\,II & 2.18 &   $-$0.76 &    0.22  &      4 &   $-$0.64 &    0.27     &   $-$0.51 &    0.18  &      2 &   $-$0.65 &    0.24 \\ 
\enddata 
\tablenotetext{a}{Values pertain to [Fe/H]}
\end{deluxetable*} 

\begin{deluxetable*}{lrrrrrrrrrrr} 
\tablecolumns{12}
\tabletypesize{\tiny}                                                                                                   
\tablewidth{0pc} 
\tablecaption{\label{Tab:xfe}1D LTE ABUNDANCES OF THE FINAL THREE BO\"{O}TES I GIANTS} 
\tablenum{6} 
\tablehead{ 
\colhead {Species} & {log\,$\epsilon_{\odot}$}& {log\,$\epsilon$}  & {s.e.$_{\log\epsilon}$}    & {N}   & {[X/Fe]} & {$\sigma$[X/Fe]} & {log\,$\epsilon$} & {s.e$_{\log\epsilon}$}     & {N}   & {[X/Fe]} & {$\sigma$[X/Fe]}\\
     ~           &      ~                  & (NY)               & (NY)      & (NY)  & (NY)     & (NY)    & (GM)              & (GM)       & (GM)  & (GM)     & (GM)\\
            ({1})  & ({2})              & ({3})     & ({4}) & ({5})    & ({6})   & ({7})             & ({8})      & ({9}) & ({10})   & ({11}) & ({12})
}
\startdata 
 & & & & & & & & & & & \\        
Boo-119 & & & & & & & & & & & \\          
Na\,I  & 6.24  &    3.73 &      ... &      1 &    0.70 &    0.20     &    3.55 &      ... &      1 &    0.75 &    0.26 \\ 
Mg\,I  & 7.60  &    5.33 &      ... &      1 &    0.94 &    0.18     &    5.30 &      ... &      1 &    1.14 &    0.25 \\ 
Si\,I  & 7.51 &     ... &     ...  &     .. &     ... &     ...     &     ... &     ...  &     .. &     ... &     ... \\ 
Ca\,I  & 6.34 &    3.39 &    0.09  &      4 &    0.26 &    0.10     &    3.57 &      ... &      1 &    0.67 &    0.25 \\ 
Sc\,II & 3.15 &     ... &     ...  &     .. &     ... &     ...     &     ... &     ...  &     .. &     ... &     ... \\ 
Ti\,I  & 4.95 &     ... &     ...  &     .. &     ... &     ...     &    2.19 &      ... &      1 &    0.69 &    0.25 \\ 
Ti\,II  & 4.95 &     ... &     ...  &     .. &     ... &     ...     &    2.43 &      ... &      1 &    0.92 &    0.30 \\ 
Cr\,I  & 5.64 &    1.94 &    0.18  &      2 &   $-$0.49 &    0.19     &    2.26 &      ... &      1 &    0.06 &    0.25 \\ 
Fe\,I\tablenotemark{a}  & 7.50  &    4.29 &    0.04  &     18 &   $-$3.21 &    0.16     &    4.06 &    0.05  &     24 &   $-$3.44 &    0.16 \\ 
Fe\,II & 7.50  &    3.51 &      ... &      1 &   $-$0.78 &    0.28     &     ... &     ...  &     .. &     ... &     ... \\ 
Ni\,I  & 6.22  &     ... &     ...  &     .. &     ... &     ...     &     ... &     ...  &     .. &     ... &     ... \\ 
Zn\,I  & 4.56 &     ... &     ...  &     .. &     ... &     ...     &     ... &     ...  &     .. &     ... &     ... \\ 
Y\,II  & 2.21 &     ... &     ...  &     .. &     ... &     ...     &     ... &     ...  &     .. &     ... &     ... \\ 
Ba\,II & 2.18 &   $-$2.03 &      ... &      1 &   $-$1.00 &    0.24     &     ... &     ...  &     .. &     ... &     ... \\ 
 & & & & & & & & & & & \\                 
Boo-127 & & & & & & & & & & & \\          
Na\,I  & 6.24  &     ... &     ...  &     .. &     ... &     ...     &     ... &     ...  &     .. &     ... &     ... \\ 
Mg\,I  & 7.60  &    5.63 &      ... &      1 &    0.14 &    0.21     &    5.88 &    0.05  &      2 &    0.20 &    0.14 \\ 
Si\,I  & 7.51 &     ... &     ...  &     .. &     ... &     ...     &    5.72 &      ... &      1 &    0.13 &    0.23 \\ 
Ca\,I  & 6.34 &    4.45 &    0.04  &     11 &    0.22 &    0.05     &    4.53 &    0.04  &     13 &    0.10 &    0.05 \\ 
Sc\,II & 3.15 &    0.99 &    0.02  &      3 &   $-$0.04 &    0.20     &    1.27 &    0.06  &      5 &    0.03 &    0.19 \\ 
Ti\,I  & 4.95 &    2.96 &    0.09  &      7 &    0.13 &    0.11     &    3.27 &    0.07  &     10 &    0.25 &    0.09 \\ 
Ti\,II & 4.95 &    3.07 &    0.04  &      4 &    0.23 &    0.18     &    3.23 &    0.06  &      7 &    0.19 &    0.19 \\ 
Cr\,I  & 5.64  &    3.40 &    0.06  &      5 &   $-$0.12 &    0.07     &    3.67 &    0.07  &      6 &   $-$0.06 &    0.08 \\ 
Fe\,I\tablenotemark{a}  & 7.50  &    5.39 &    0.03  &     59 &   $-$2.11 &    0.16     &    5.58 &    0.02  &     64 &   $-$1.92 &    0.16 \\ 
Fe\,II & 7.50  &    5.41 &    0.09  &      3 &    0.03 &    0.24     &    5.72 &    0.09  &      4 &    0.13 &    0.23 \\ 
Ni\,I  & 6.22  &    4.12 &    0.09  &      2 &    0.02 &    0.15     &    4.16 &    0.10  &     11 &   $-$0.15 &    0.11 \\ 
Zn\,I  & 4.56 &     ... &     ...  &     .. &     ... &     ...     &     ... &     ...  &     .. &     ... &     ... \\ 
Y\,II  & 2.21 &     ... &     ...  &     .. &     ... &     ...     &     ... &     ...  &     .. &     ... &     ... \\ 
Ba\,II & 2.18 &   $-$0.49 &    0.25  &      4 &   $-$0.56 &    0.30     &   $-$0.55 &    0.01  &      2 &   $-$0.81 &    0.28 \\ 
 & & & & & & & & & & & \\                 
Boo-130 & & & & & & & & & & & \\          
Na\,I  & 6.24  &    3.92 &      ... &      1 &    0.03 &    0.21     &     ... &     ...  &     .. &     ... &     ... \\ 
Mg\,I  & 7.60 &    5.35 &      ... &      1 &    0.10 &    0.19     &    5.64 &      ... &      1 &    0.32 &    0.24 \\ 
Si\,I  & 7.51 &     ... &     ...  &     .. &     ... &     ...     &     ... &     ...  &     .. &     ... &     ... \\ 
Ca\,I  & 6.34 &    4.06 &    0.08  &      6 &    0.07 &    0.09     &    4.37 &    0.05  &     11 &    0.31 &    0.06 \\ 
Sc\,II & 3.15 &    0.73 &      ... &      1 &   $-$0.07 &    0.28     &    0.89 &    0.04  &      5 &    0.03 &    0.19 \\ 
Ti\,I  & 4.95 &    2.60 &    0.08  &      3 &    0.01 &    0.13     &    2.79 &    0.08  &      7 &    0.13 &    0.10 \\ 
Ti\,II & 4.95 &    2.94 &    0.15  &      3 &    0.34 &    0.23     &    2.89 &    0.02  &      6 &    0.22 &    0.18 \\ 
Cr\,I  & 5.64 &    3.28 &    0.25  &      4 &   $-$0.01 &    0.25     &    3.49 &    0.06  &      5 &    0.13 &    0.07 \\ 
Fe\,I\tablenotemark{a}  & 7.50  &    5.15 &    0.03  &     55 &   $-$2.35 &    0.16     &    5.22 &    0.04  &     35 &   $-$2.28 &    0.16 \\ 
Fe\,II & 7.50  &    5.16 &    0.10  &      2 &    0.01 &    0.25     &    5.45 &    0.11  &      3 &    0.24 &    0.25 \\ 
Ni\,I  & 6.22  &    3.85 &    0.14  &      3 &   $-$0.02 &    0.15     &    4.06 &    0.06  &      5 &    0.12 &    0.07 \\ 
Zn\,I  & 4.56 &    2.47 &      ... &      1 &    0.26 &    0.24     &     ... &     ...  &     .. &     ... &     ... \\ 
Y\,II  & 2.21 &     ... &     ...  &     .. &     ... &     ...     &     ... &     ...  &     .. &     ... &     ... \\ 
Ba\,II & 2.18 &   $-$0.74 &    0.23  &      3 &   $-$0.57 &    0.28     &   $-$0.61 &    0.06  &      2 &   $-$0.51 &    0.19 \\ 
\enddata 
\tablenotetext{a}{Values pertain to [Fe/H]}
\end{deluxetable*}

%% file: tab7.tex
\begin{deluxetable}{lrrrr}                                                                                                                  
\tablecolumns{5}                                                                                                                            
\tablewidth{0pt}                                                                                                                  
\tablecaption{\label{Tab:errors}ABUNDANCE UNCERTAINTIES IN [X/Fe]}                   
\tablenum{7} 
\tablehead{                                                                                                                              
  \colhead{Species} & {$\Delta${\teff}} & {$\Delta${\logg}} & {$\Delta\xi_t$} & {$\Delta$[X/Fe]} \\   
             {}     & {(100K)}          & {(0.3~dex)}       & {(0.3~{\kms})}  &                  \\        
           {(1)}    & {(2)}             & {(3)}             &{(4)}            & {(5)}                                   
  }                                                                                                                                       
\startdata                                                                                                                               

Na\,I  &    0.040   &   $-$0.042 &   $-$0.060 &    0.083 \\
Mg\,I  &   $-$0.035 &   $-$0.018 &    0.000   &    0.039 \\
Si\,I  &   $-$0.080 &    0.036   &    0.105   &    0.137 \\
Ca\,I  &   $-$0.025 &    0.003   &    0.030   &    0.039 \\
Sc\,II &   $-$0.115 &    0.135   &    0.045   &    0.183 \\
Ti\,I  &    0.050   &   $-$0.009 &    0.030   &    0.059 \\
Ti\,II &   $-$0.115 &    0.135   &    0.015   &    0.178 \\
Cr\,I  &    0.040   &   $-$0.006 &    0.015   &    0.043 \\
Fe\,I\tablenotemark{a}  &    0.125   &   $-$0.027 &   $-$0.090 &    0.156 \\
Fe\,II &   $-$0.150 &    0.144   &    0.045   &    0.213 \\
Ni\,I  &    0.005   &    0.012   &    0.045   &    0.047 \\
Zn\,I  &   $-$0.110 &    0.090   &    0.045   &    0.149 \\
Y\,II  &   $-$0.100 &    0.138   &    0.045   &    0.176 \\
Ba\,II &   $-$0.085 &    0.135   &   $-$0.015 &    0.160 
\enddata                                                                                                      
\tablenotetext{a}{Errors pertain to uncertainties in [Fe/H]}
\end{deluxetable}

%% file: tab8.tex
\begin{deluxetable*}{lccccrccccc} 
\tablecolumns{11} 
\tablewidth{0pt}  
\tablecaption{\label{Tab:alphas}ADOPTED $\alpha-$ELEMENT ABUNDANCES}                   
\tablenum{8} 
\tablehead{ 
\colhead{Star} & [Mg/Fe] & $\sigma$\tablenotemark{b} & [Ca/Fe] &  $\sigma$\tablenotemark{b} & [Ti/Fe] & $\sigma$\tablenotemark{b} & [$\alpha$/Fe]\tablenotemark{a} & $\sigma$\tablenotemark{c} & [Fe/H] &  $\sigma$\tablenotemark{b} \\    
  }
%                           mg             ca              ti            alfa
\startdata 
Boo-33     &                 0.26 & 0.22 & 0.14 & 0.06 & $-$0.02 & 0.17 & 0.13 & 0.17 & $-$2.32 & 0.16 \\
Boo-41     &                 0.50 & 0.22 & 0.28 & 0.06 &   0.78  & 0.20 & 0.52 & 0.18 & $-$1.88 & 0.16 \\
Boo-94     &                 0.49 & 0.16 & 0.30 & 0.06 &   0.26  & 0.13 & 0.35 & 0.10 & $-$2.94 & 0.16 \\
Boo-117    &                 0.18 & 0.22 & 0.20 & 0.06 &   0.14  & 0.13 & 0.18 & 0.15 & $-$2.18 & 0.16 \\
Boo-119\tablenotemark{d} &   1.04 & 0.22 & 0.46 & 0.18 &   0.80  & 0.28 & 0.77 & 0.15 & $-$3.33 & 0.16 \\
Boo-127    &                 0.17 & 0.18 & 0.16 & 0.05 &   0.20  & 0.14 & 0.18 & 0.13 & $-$2.01 & 0.16 \\
Boo-130    &                 0.21 & 0.22 & 0.19 & 0.08 &   0.17  & 0.16 & 0.19 & 0.16 & $-$2.32 & 0.16 \\
Boo-1137   &                 0.30 & 0.21 & 0.55 & 0.14 &   0.48  & 0.10 & 0.44 & 0.09 & $-$3.66 & 0.11 
\enddata      
\tablenotetext{a}{[$\alpha$/Fe] = ([Mg/Fe] + [Ca/Fe] + [Ti/Fe])/3 }
\tablenotetext{b}{Arithmetic mean of NY and GM errors}
\tablenotetext{c}{Quadrature mean of NY and GM errors, except for
  Boo-119 and Boo-1137 for which the values are based on the analysis
  of GM and the present reanalysis of the data of \citet{norris10a},
  respectively}
\tablenotetext{d}{CEMP-no star
\vspace{3mm} 
}
\end{deluxetable*}

%% file: tab9.tex
\begin{deluxetable*}{lcccccccccccc}                                                                                                                  
\tablecolumns{13}                                                                                                                            
\tabletypesize{\scriptsize}
\tablewidth{0pt}                                                                                                                  
\tablecaption{\label{Tab:dispersions}RELATIVE-ABUNDANCE DISPERSIONS}  
\tablenum{9} 
\tablehead{                                                                                                                              
\colhead {Species} & 
{$\sigma_{\rm Boo}$} & 
{$\sigma_{\rm meas}$} & 
{N$_{\rm stars}$\tablenotemark{a}}   & 
{$\langle$N$_{\rm lines}$$\rangle$} & 
{$\sigma_{\rm Boo}$} & 
{$\sigma_{\rm meas}$} & 
{N$_{\rm stars}$\tablenotemark{a}  }& 
{$\langle$N$_{\rm lines}$$\rangle$} & 
{$\sigma_{\rm Halo}$} & 
{$\sigma_{\rm meas }$\tablenotemark{b}} & 
{N$_{\rm stars}$}   & 
{$\langle$N$_{\rm lines}$$\rangle$} \\   
            {}     &     
{NY}   &    
{NY}    &    
{NY}    & 
{NY}     & 
{ GM}      & 
{GM}       &   
{GM  }  & 
{GM}     & 
{C04}          & 
{C04}& 
{C04}          & 
{C04}           \\   
            {(1)}  &  
{(2)}     & 
{(3)}      & 
{(4)}      & 
{(5)}    & 
{(6)}      & 
{(7)}      & 
{(8)}     & 
{(9)}    & 
{(10)}     & 
{(11)} & 
{(10)}     & 
{(11)} 
}                                                                                                                                       
\startdata                                                                                                    
$[$Ca\,I/Fe$]$       &   0.11    &   0.07    &  6         &     8.8  &   0.08    &   0.06    &  6        &    10.3  &   0.06    &   0.07 & 13 & 15.8 \\ 
$[$Ti\,I/Fe$]$       &   0.31    &   0.11    &  6         &     4.8  &   0.27    &   0.09    &  6        &     7.8  &   0.06    &   0.05 & 13 & 12.9 \\ 
$[$Cr\,I/Fe $]$      &   0.11    &   0.11    &  6         &     4.8  &   0.21    &   0.07    &  6        &     4.7  &   0.08    &   0.07 & 13 &  6.9 \\ 
$[$Ni\,I/Fe $]$      &   0.12    &   0.17    &  6         &     2.5  &   0.12    &   0.08    &  6        &     6.2  &   0.09    &   0.09 & 13 &  3.0  
\enddata                                                                                                      
\tablenotetext{a}{In all cases, the six stars are Boo-33, Boo-41, Boo-94, Boo-117, Boo-127, and Boo-130}
\tablenotetext{b}{From \citet[their Table~9, column (7)]{cayrel04}}
\end{deluxetable*}